\documentclass{article}
\usepackage{lmodern}
\usepackage[T1]{fontenc}
\usepackage[latin9]{inputenc}
\usepackage{geometry}
\usepackage{color}
\usepackage{amsmath}
\usepackage{amssymb}
\usepackage{stmaryrd}
\usepackage{graphicx}
\usepackage{esint}
\usepackage{wrapfig}
\usepackage[unicode=true,pdfusetitle,
 bookmarks=true,bookmarksnumbered=false,bookmarksopen=false,
 breaklinks=false,pdfborder={0 0 1},backref=false,colorlinks=true]
 {hyperref}

\makeatletter

\newcommand{\lyxdot}{.}

\usepackage[mathletters]{ucs}
\usepackage[utf8x]{inputenx}
\usepackage{amsthm}\usepackage{xspace}\usepackage[all,cmtip]{xy}
\usepackage{comment}
\usepackage[vcentermath]{youngtab}
\usepackage{pdfpages}

\newtheorem{cha}{Problem}
\makeatother

\begin{document}
\global\long\def\rbracket#1{\left|#1\right\rangle }
\global\long\def\lbracket#1{\left\langle #1\right|}

\title{Obituary for a Flea\footnote{The research in this paper is part of the project \emph{Experimental Tests of Quantum Reality}, funded by the \emph{Templeton World Charity Foundation}}}
 \author{Jasper van Heugten\footnote{Radboud Universiteit Nijmegen, Institute for Mathematics, Astrophysics, and Particle Physics, Heyendaalseweg
135, 6525 AJ Nijmegen, The Netherlands.}\ \ and Sander Wolters\footnote{Radboud Universiteit Nijmegen, Institute for Mathematics, Astrophysics, and Particle Physics, Heyendaalseweg
135, 6525 AJ Nijmegen, The Netherlands. E-mail: sretlowrednas@gmail.com}}
\maketitle
\begin{abstract}
The Landsman-Reuvers proposal to solve the measurement problem from within quantum theory is extensively analysed. In favor of proposals of this kind, it is shown that the standard reasoning behind objections to solving the measurement problem from within quantum theory rely on counterfactual reasoning or mathematical idealisations. Subsequently, a list of objections/challenges to the proposal are made. Part of these objections are equally important for all attempts at solving the measurement problem, such as the problem of interpreting small numbers in the density matrix, the problem of reproducing the Born rule, the use of pure states as a tool to alleviate the interpretational issues of quantum states, and the necessity of introducing classical certainties which are not strictly present in quantum theory. The additional objections that are particular to the proposal, such as the physical interpretation/origin of the flea perturbation, the use of potentials to solve a dynamical problem, slow collapse times, the inability to handle unequal probabilities, and the dictatorial role of the flea perturbation, lead us to believe that the Landsman-Reuvers proposal is lacking in both physical grounding and theoretical promise. Finally, an overview is given of the challenges that were encountered in this attempt to solve the measurement problem from within quantum theory.
\end{abstract}

\section{Overview}

In 2013, Landsman, and his student Reuvers, proposed a reformulation of the measurement problem of quantum mechanics~\cite{lare}. Using the sensitivity of bound states with respect to small perturbations in the semi-classical limit, it was claimed that this reformulation may hold the possibility of resolving the measurement problem within the formalism of quantum theory. We shall refer to this idea and the (simple) mathematical model expressing this idea as the \emph{flea model} or \emph{flea approach}. Later~\cite{landsman}, Landsman embedded his treatment of the measurement problem within a larger framework regarding, but not limited to, the relations between classical and quantum physics. Following the contribution of Landsman and Lindenhovius in this volume, we refer to this framework as \emph{asymptotic Bohrification}. \\

Our goals are twofold:
\begin{enumerate}
\item To investigate the feasibility of the flea model.
\item To analyse to what extent asymptotic Bohrification captures the measurement problem.
\end{enumerate}

The bulk of the text below is divided into three parts. In Section~\ref{sec:mp} we recall the measurement problem, with emphasis on the problem of outcomes. Considering the von Neumann measurement model, and generalisations thereof, we argue that in order to arrive at the conclusion that quantum theory cannot account for the occurrence of outcomes, we need counterfactual reasoning or mathematical idealisations. Removing these idealisations opens the possibility, at least in principle, of solving the problem of outcomes within quantum theory. On the other hand, removing these idealisations adds subtlety to the formulation of the measurement problem. In particular, we require post-measurement outcome states for which it is unclear whether we can comprehend these as carriers of properties of the measured system, or the pointer.

This brings us to Section~\ref{sec:bohr} where we first consider the reformulation of the problem of outcomes within asymptotic Bohrification. This framework aims to avoid letting mathematical idealisations play a decisive role, and, as a consequence, it has to deal with the aforementioned post-measurement states. These are states where the probabilities assigned to the different pointer values, by the Born rule,  are all non-zero. Mathematically, asymptotic Bohrification bridges the gap between these approximate outcome states and the conceptually clear classical outcomes. However, regardless of how suggestive the mathematics of the semi-classical limit might appear, in order to bridge this gap physically, we need to a priori assume the Born rule. This situation is reminiscent of the objections of circularity in the derivation of the Born rule in everettian quantum mechanics, and of the tails problem for the GRW models for spontaneous collapse. 

But should we be reluctant in a priori assuming the Born rule? Solving the problem of outcomes seems to be a considerable challenge, even with this added assumption. Our reluctance with respect to assuming the Born rule is explained as follows. It is not just the post-measurement state which is idealised (through assumptions such as the eigenstate-eigenvalue link) but also the initial state. This state is typically assumed to be factorised between the system and the environment (i.e, all other degrees of freedom which we consider relevant). This ignores entanglement of the system with the environment and the fundamental uncertainties of quantum physics. Such idealisations, motivated by the Born rule, should be embraced by the pragmatist working with the formalism of quantum theory, since such assumptions barely affect the statistics of the outcomes. However, in studying the problem of outcomes, and hopefully the origin of the Born rule, these idealisations may amount to throwing away essential pieces of information. Even though the factorised states or pure states that are typically used in calculations, are close to the actual states in experiments in a precise mathematical sense and in terms of statistics of outcomes, they are conceptually very different due to entanglement with the environment. 

One possible indication in that direction is the Problem of independence, discussed in Subsection~\ref{sub:born_baby_born} and re-evaluated for the flea proposal in Subsection~\ref{sub:dis3}. If we only allow for unitary time-evolution (when all relevant degrees of freedom are included), fixing the initial state and the hamiltonian determines the outcome state. To obtain different outcomes for a single-system preparation method we need to slightly vary the initial environmental state and/or the dynamics. However, the statistics obtained by these variations must coincide with the probabilities obtained by the Born rule from the reduced system state. The \emph{problem of independence} is the explanatory gap relating the Born probabilities to the statistics of variations in dynamics. This problem is particularly pressing for the flea approach where the variations in dynamics bear no relation to the system state. If we believe the problem of outcomes to be solvable within quantum physics, then taking entanglement of the initial system state with the environment into account may turn out to be essential in solving the problem of independence.

We do not necessarily view the problems of understanding the use of approximate initial and final states and the problem of independence as a weakness of asymptotic Bohrification. Rather, we see it as inevitable challenges for any approach to the problem of outcomes within the formalism of quantum theory. Of course, it may turn out that the problem of outcomes cannot be solved within the confines of quantum theory. As noted before, it is not clear whether we can think of these non-idealised initial and final reduced states as a collection of properties of the system. Possibly we are relying on a erroneous ontology of states, and only ensemble interpretations are to be allowed. Regardless, at this point it is speculative whether or not the problem of outcomes can be solved within quantum theory, and we deem this question important enough to warrant further research.\\

In Section~\ref{sec:flea} we concentrate on the flea model. Ideally, studying this model would provide hints on how to attack the problem of independence. However, such an approach is hindered by the fact that the environment does not play an explicit role in the flea model. Consequently, it is not even clear whether or not the used flea perturbation is of purely quantum mechanical origin. In addition, unrealistically long collapse times and problems in generalising the model lead us to conclude that the flea model is not feasible as a model for dynamical collapse of the wave function. When applying the flea approach to existing models of quantum measurement \cite{abn,hasp}, we are thwarted by the problem of independence. In our attempt to apply the flea proposal, the flea acts only at the level of the pointer variable, this implies, however, that if the perturbation is effective, that it removes the correlation between the pointer and the observable. In other words, we no longer have a measurement. As was recommended to us by Landsman, the flea should act on the combined pointer and observable to solve this problem. We briefly discuss the difficulties in understanding this recommendation to apply the flea proposal in a realistic model on the level of pointer and observable.\\

It is still unclear whether the problem of outcomes can be solved within quantum physics. What is clear, is that the flea approach is unfeasible as a solution, and the work of Jona-Lasinio et.al. \cite{jola} upon which it draws its inspiration need not even be relevant to a dynamical collapse. If we are to find a solution for the problem of outcomes within quantum theory, like the one formulated within asymptotic Bohrification, then the problem of independence needs to be addressed first. First, one needs to demonstrate that variations in dynamics, serving the function of the flea perturbation, can actually arise from the environment, and how the details of such a variation depends on the preparation of the system. Only then is it clear whether a "flea" can be of quantum mechanical origin, and whether the proposed solution to the problem of outcomes can be consistent with the Born rule. And only then can we effectively attack the problem of how to tweak models of quantum measurement accordingly to investigate this further.

\section{The Measurement Problem} \label{sec:mp}

Ever since the relevance of environment-induced decoherence became acknowledged, it is customary to divide the measurement problem into three subproblems. We briefly consider this in Subsection~\ref{sub:sub}, after which we concentrate solely on the subproblem of outcomes. In Subsections~\ref{sub:poo2} and \ref{sub:poo1} we consider different versions of von Neumann's model for an ideal measurement. For these models the impossibility for quantum physics in assigning outcomes with respect to arbitrary initial system states depends crucially on mathematical idealisations, or, alternatively, on counterfactual reasoning. Dismissing such idealisations opens the possibility, at least in principle, of solving the problem of outcomes. Exploiting this loophole comes at the price of added subtlety. The post-measurement states are in need of clarification in order to be understood as outcomes, and there is a conceptual gap where we need to connect variations in initial conditions and/or dynamics to the Born probabilities extracted from the system state. In Subsection~\ref{sub:uitkom} it is discussed how pure states, or density matrices with some zeros on the diagonal, are often used as outcome states to insert certainties into models that are not strictly present in quantum theory. Then, in Subsection~\ref{sub:epsi}, a common mistake in interpreting approximate outcomes is analyzed. In the next section we consider the reformulation of the problem of outcomes from the perspective of asymptotic Bohrification, an approach which openly dismisses physical arguments which crucially hinge on the use of mathematical idealisations. The subtleties of the current section then reappear as challenges to asymptotic Bohrification.

\subsection{Subproblems} \label{sub:sub}

As is customary, we divide the measurement problem into three subproblems. The first of these problems is the \emph{preferred basis problem}, and it is somewhat tricky to give a brief yet precise account which completely captures this problem. Therefore, we try to explain it at the hand of an example. Consider the following (simplified) version of an ideal measurement. Suppose that we are interested in measuring a (non-degenerate) observable associated to a two-level system $\mathcal{S}$. Let $\vert m_{1}\rangle$, $\vert m_{2}\rangle$ be an orthonormal basis for the Hilbert space $\mathcal{H}_{\mathcal{S}}$, consisting of eigenstates of the observable. Let $\vert r\rangle$ be a suitable ready state for the experiment, representing the initial state of the pointer variable of the measuring device, and any other degree of freedom which we deem relevant in the measurement interaction. The initial states of the form $\vert m_{i}\rangle\otimes\vert r\rangle$, with $i\in\{1,2\}$, are assumed to evolve into states of the form $\vert m_{i}\rangle\otimes\vert E_{i}\rangle$, where $\vert E_{i}\rangle$ is an environmental state to which we can assign a value $x_{i}$ to the pointer variable. For a well-designed measurement the values $x_{1}$ and $x_{2}$ are macroscopically distinct. We take this into account by assuming the orthogonality condition\footnote{We take a very strict condition for two states to be distinct for illustration purposes. Although macroscopic distinctiveness is hard to define, we at least expect the overlap to become close to zero in some limit of macroscopicity. The arguments below will then be approximate, but the problem remains of defining a physical mechanism which chooses an (approximate) basis in which off-diagonal elements will tend to disappear (problem of unobservability of interference) and/or the selection of one of the outcomes in this basis will occur (problem of outcomes), depending on the problem we wish to solve.} $\langle E_{1}\vert E_{2}\rangle=0$.

Next, consider the initial state
\begin{equation*}
\vert\phi\rangle\otimes\vert r\rangle,\ \ \ \vert\phi\rangle:=\frac{1}{\sqrt{2}}\left(\vert m_{1}\rangle+\vert m_{2}\rangle\right).
\end{equation*}
By linearity of the Schr\"odinger equation, over time this state becomes entangled with the environment as
\begin{equation*}
\vert\Phi\rangle=\frac{1}{\sqrt{2}}\left(\vert m_{1}\rangle\otimes\vert  E_{1}\rangle+\vert m_{2}\rangle\otimes\vert E_{2}\rangle\right).
\end{equation*}

Looking only at the post-measurement state $\vert\Phi\rangle$, we are unable to determine the measurement basis (i.e., the eigenstates of the observable, or, equivalently, the observable itself). With respect to the basis $\vert m_{1}\rangle$, $\vert m_{2}\rangle$, the reduced system state, after tracing out all environmental degrees of freedom, is (of course) diagonal. However, consider the alternative orthonormal basis
\begin{equation*}
\vert m_{+}\rangle=\frac{1}{\sqrt{2}}\left(\vert m_{1}\rangle+\vert m_{2}\rangle\right),\ \ \vert m_{-}\rangle=\frac{1}{\sqrt{2}}\left(\vert m_{1}\rangle-\vert m_{2}\rangle\right),
\end{equation*}
corresponding to an observable which does not commute with our original observable. With respect to the orthonormal basis, $\vert m_{+}\rangle$, $\vert m_{-}\rangle$, the reduced system state is diagonal as well. There is an infinity of pairwise non-commuting observables yielding bases with respect to which the reduced system state is diagonal. In itself, the fact that we are unable to pinpoint the exact measurement basis from a single post-measurement state is not remarkable. Why should it reveal the measurement basis or observable? In addition, we already knew the observable from the way the model was set up. The interaction between system and environment was chosen as to let $\vert m_{i}\rangle\otimes\vert r\rangle$ evolve into $\vert m_{i}\rangle\otimes\vert E_{i}\rangle$. In other words, restricted to suitable initial states, the time evolution operator, expressed as a $2\times2$ matrix of operators acting on the Hilbert space $\mathcal{H}_{\mathcal{E}}$, becomes diagonal with respect to the $\vert m_{i}\rangle$ basis. By construction, the measurement basis $\vert m_{1}\rangle$, $\vert m_{2}\rangle$ was defined to be a basis of vectors which do not get entangled with the environment. 

The preferred basis problem is the problem that for any model which represents a measurement performed on a quantum system, and possibly an even larger class of examples falling within the scope of the quantum to classical transition, we need a method for choosing a \emph{unique} basis with respect to which the reduced system state becomes diagonal. Environment-induced decoherence provides a wide range of examples, typically featuring a macroscopic environment, in which a unique basis is selected, with respect to which the basis states become minimally entangled with the environment. With respect to this basis, the off-diagonal elements of the reduced system state become extremely small, on a calculable short time scale. Note, however, that the system state is not exactly diagonal with respect to the environment selected basis.\\

The smallness of the off-diagonal terms of the reduced system state, with respect to the decoherence selected basis, is also relevant to the second subproblem of the measurement problem, the \emph{problem of unobservability of interference}. Interference effects, such as in the infamous double slit experiment for electrons, are ubiquitous in quantum mechanics. Yet, if we consider mesoscopic or macroscopic objects, it becomes extremely hard to achieve experimental conditions which provide interference patterns. Environment-induced decoherence explains the general absence of interference phenomena, and, tied to that, the absence of superpositions of macroscopically distinguishable states. With respect to the decoherence selected basis, the  reduced system state is approximately diagonal. The reduced state approximates that of an ensemble of basis states much more than it does a superposition. Of course, these superpositions of basis states are typically there. But because of entanglement with the environment, when we concentrate only on the system by tracing out this environment, we lose these superpositions.\\

The third, and final, subproblem is the \emph{problem of outcomes}. How can quantum mechanics account for the observation that measurements have outcomes? Clearly, to what extent this is a problem depends heavily on the philosophical stance which we take. Environment-induced decoherence, which provides us with an approximately diagonal matrix, but not a way to select a single entry on the diagonal, has not provided an answer to this problem. We consider the problem of outcomes in detail in the rest of this section.

\subsection{Problem of Outcomes I} \label{sub:poo2}

The first version of the measurement problem was taken from~\cite{bagi} and is of interest because of its generality and because it uses a sufficiently liberal notion of outcome, making it relevant to the flea model. As before, we use a two-level system $\mathcal{S}$. Let $\vert +\rangle$ and $\vert-\rangle$ be an orthonormal basis of the Hilbert space $\mathcal{H}_{\mathcal{S}}$. The post-measurement states are written in the form $\vert A,\alpha\rangle$. The label $A$ denotes a value of a certain macroscopic variable, which we call the pointer variable. With respect to the state $\vert A,\alpha\rangle$, the pointer is assigned the value $A$. The second index, $\alpha$, refers to all other degrees of freedom we deem relevant. These might belong to the system $\mathcal{S}$, or be environmental, and may even include the whole universe (apart from the pointer variable). 

Let $V_{A}$ denote the set of all states $\vert A,\alpha\rangle$, for which we agree that it is sensible to say that the pointer variable has value $A$. Suppose that $A$ and $B$ are two macroscopically distinguishable values for the pointer variable. We shall not assume that the sets $V_{A}$ and $V_{B}$ are closed linear subspaces of the relevant Hilbert space, or, even stronger, that there are associated projections which are orthogonal. We consider a weaker relation between the elements of $V_{A}$ and $V_{B}$ instead. This weaker relation allows for the possibility of states in which the pointer has an outcome, but which display some form of tail with respect to the pointer variable. Recall that if $\vert\Phi\rangle$ and $\vert\Psi\rangle$ are orthogonal vectors, then $\Vert\vert\Phi\rangle-\vert\psi\rangle\Vert^{2}=2$. We assume that elements of $V_{A}$ and $V_{B}$ are close to orthogonal in the sense that there exists a number $\eta\ll1$ such that for all $\vert A,\alpha\rangle$ in $V_{A}$ and all $\vert B,\beta\rangle$ in $V_{B}$ we have,
\begin{equation} \label{eq_almostortho}
\Vert\vert A,\alpha\rangle-\vert B,\beta\rangle\Vert\geq\sqrt{2}-\eta.
\end{equation}
Alternatively, this assumption follows from assuming that for macroscopically distinguishable pointer values $A$ and $B$, there exists a positive number $\epsilon\ll1$, such that the transition probability satisfies
\begin{equation}
p(\vert A,\alpha\rangle\ \vert\ \vert B,\beta\rangle)=\vert\langle A,\alpha\vert B,\beta\rangle\vert\leq\epsilon,
\end{equation}
for all states in $V_{A}$ paired with states in $V_{B}$. The initial pre-measurement state is assumed to be factorised
\begin{equation*}
\left(a\vert+\rangle+b\vert-\rangle\right)\otimes\vert\Phi\rangle,
\end{equation*}
where $\vert\Phi\rangle$ is an appropriate initial state of the measurement apparatus and the environment. We do not need to further specify this state. The post-measurement state is not assumed to be factorised. Time evolution is assumed to be given by a unitary transformation, and is in particular linear. An initial state of the form $\vert+\rangle\otimes\vert\Phi\rangle$ evolves to a state $\vert P,\alpha\rangle$, with pointer value $P$. In~\cite{bagi} it is only assumed that the final state is of the form $\vert P,\alpha\rangle$ with a high probability. For some choices of $\vert\Phi\rangle$ the post-measurement state need not assign the value $P$ to the pointer. For our current purposes these probabilistic considerations play no role and we shall not consider them any further. We assume that initial states of the form $\vert-\rangle\otimes\vert\Phi\rangle$ evolve into states $\vert M,\beta\rangle$ for which we attribute the value $M$ to the pointer.

The measurement problem arises when the initial system state is taken to be
\begin{equation} \label{eq_spi}
\vert\psi\rangle=\frac{1}{\sqrt{2}}\left(\vert+\rangle+\vert-\rangle\right).
\end{equation}
By linearity the measurement interaction yields the evolved state
\begin{equation} \label{eq_finalstate}
\vert\psi\rangle\otimes\vert\Phi\rangle\mapsto U(t_{0},t_{f})\vert\psi\rangle\otimes\vert\Phi\rangle=\frac{1}{\sqrt{2}}\left(\vert P,\alpha\rangle+\vert M,\beta\rangle\right).
\end{equation}
We assume that the pointer values $M$ and $P$ are macroscopically distinguishable. As a result $\vert P,\alpha\rangle\in V_{P}$ and $\vert M,\beta\rangle\in V_{M}$ are almost orthogonal in the sense of the inequality (\ref{eq_almostortho}). But from (\ref{eq_finalstate}) we deduce that with respect to the norm metric, the distance of the evolved state to both $V_{P}$ and $V_{M}$ is at most 1. As a consequence, the post-measurement state is neither an element of $V_{P}$, nor an element of $V_{M}$, and hence we are unable to assign any outcome to the evolved state.\\

This formulation of the problem of outcomes does not depend on any idealisations, at least at first sight. In anticipation of the next section, the usage of a pure initial state may itself be seen as an idealisation, but this has no impact on the issue at hand. So, at first sight the reasoning seems to provide a convincing argument that quantum mechanics is ill equipped for explaining the fact that measurements yield outcomes. This is an argument which presents a strong case that quantum theory needs to be modified (the strategy pursued in~\cite{bagi}) or a new understanding of the formalism is needed (such as in everettian quantum mechanics) in order to resolve the problem of outcomes. However, in addition to the linearity of the time-evolution and the almost orthogonality of macroscopically distinguishable states, there is another (implicit) assumption which is crucial to the problem of outcomes. It is this implicit assumption which we seek to challenge. 

Given a \emph{fixed} environmental/apparatus state $\vert\Phi\rangle$ we consider different initial states $\vert\phi\rangle\otimes\vert\Phi\rangle$ of the measurement, where $\phi\in\{+,-,\psi\}$ with $\vert\psi\rangle$ defined by (\ref{eq_spi}). These initial states are then subjected to the \emph{same} unitary transformation. This looks like counterfactual reasoning. What if we consider $\vert\phi\rangle=\vert\psi\rangle$ instead of $\vert\phi\rangle=\vert+\rangle$? Applying counter-factual reasoning to quantum theory in order to understand its foundations strikes us as a potential source of mistakes. It is preferable to avoid such reasoning in any form.

So instead of a counterfactual argument, let us think of the evolution of the three different initial states $\vert\phi\rangle\otimes\vert\Phi\rangle$, where $\phi\in\{+,-,\psi\}$, as three different runs of an experiment. From this perspective,we need to justify the use of the same state $\vert\Phi\rangle$ and the same unitary transformation $U(t_{0},t_{f})$ for all three runs, since this is an idealisation. If the index $\alpha$ refers to the whole universe (except for the pointer), then for different runs we should expect the state $\vert\Phi\rangle$ to be different. However, many of these differences are thought of to be irrelevant to the measurement process. Let us trace out these degrees of freedom. Now it makes more sense to assume that $\vert\Phi\rangle$ is the same for different runs, but can we argue that reduced dynamics are the same for the different runs?

Should we care about the idealisation involved in using the same dynamics? In other words, does this idealisation have an impact on the conclusion that we drew in the problem of outcomes? Can small variations in the time evolution operator really be relevant to the problem of outcomes? Landsman and Reuvers seem to think so, as their flea proposal~\cite{lare} takes a shot at the measurement problem by employing small variations of the potential part of the hamiltonian in order to create a dynamic collapse. We explain this idea and its limitations in Section~\ref{sec:flea}.  In a similar vein, van Wezel~\cite{iamweasel} adds a small non-hermitian term to the hamiltonian in order to create a dynamical collapse. This kind of research hints that it may be worthwhile to further study the loophole of small variations in dynamics and/or initial states for the problem of outcomes.

Note that in investigating the loophole, we are not claiming that the dynamics of quantum theory needs to be modified in a non-linear way. Quantum theory is seen as fundamental and time-evolution is linear in this theory. However, in the situations where a quantum system is probed by a macroscopic device, we believe some of the details of this device and the environment to be essential in the creation of an outcome. Since these details vary for different runs, this places us in a situation where the linearity of the underlying dynamics cannot be used when comparing different runs. 

To summarise, the general form of the problem of outcomes in this subsection either needs counterfactual reasoning, or an idealisation with respect to the time-evolution operator in order to arrive at its conclusion. Dismissing counterfactual arguments, the problem of outcomes gains a loophole where small differences in the unitary operator may play a crucial role in ensuring that each measurement run gets assigned an outcome. But we are not out of the woods yet. Not by a longshot! The following subsection demonstrates that more is needed than small variations in dynamics, in order to enable a dynamical collapse of the  wave function. In addition, for the previously mentioned collapse in the harmonic crystal model of van Wezel \cite{iamweasel}, it has not been demonstrated whether the non-unitarity of the time-evolution can arise effectively from an open quantum system. That is, we do not know whether this model operates within unaltered quantum physics. The same holds for the flea model, which faces additional problems. Problems, leading us to conclude in Section~\ref{sec:flea} that it is not feasible as a model for dynamic collapse. These remarks aside, it is still an open question whether the problem of outcomes can be solved within quantum theory.

\subsection{The Born Rule} \label{sub:born_baby_born}

Stir long enough in the quagmire that is the foundations of quantum physics, and the Born rule comes floating to the top. If we are to exploit the loophole described in the previous subsection, then the Born rule presents itself as an obstacle. Assume that for a fixed initial system state, by slightly perturbing the hamiltonian or time-evolution operator for the measurement interaction, different outcomes are obtained. Each suitable perturbation then determines a corresponding outcome.

What could be the physical origin of the perturbation? For the example of the flea model, not much has been written about what could cause the flea perturbation. However, already in the abstract of \cite{lare} it is suggested that the environment is to be the source of the flea. How this is actually realised in the model is unknown, but that is not important at this point. In the flea model it is the location of the flea perturbation that determines the outcome of a measurement run. In other words, the details of the perturbation arising from the environment determine the outcome of the measurement. To be fair, the flea model does not consider arbitrary states for the two-level system $\mathcal{S}$, but only states where both outcomes are equally probable. Still, if we seek to straightforwardly extend this model to arbitrary superpositions (as in Subsection \ref{sub:genfl}), then the flea perturbation still determines the outcome all by itself.

When first getting acquainted with quantum physics, it is easy to make the mistake of being a naive realist in the sense of believing that a measurement simply reveals a property of a quantum system. In reality, both the system on the one hand, and the apparatus plus environment on the other play an active role in creating an outcome. If we think that an environmentally induced perturbation completely determines the outcome of an experiment, aren't we committing ourselves to a position which is just as extreme as that of the naive realist? The difference being that instead of ignoring everything apart from the system $\mathcal{S}$, we are now ignoring the system itself. 

It is the Born rule which tells us how the relative frequencies of outcomes are related to the initial state vector of the system. The statistics of the perturbations determine the relative frequencies of the outcomes. As such, these environmental differences must be related to the initial system state, and the Born role must arise in a natural way. Of course, we may choose to ignore the Born rule and attempt the  challenge of finding small perturbations which provide a suitable model for a collapse on a single run without (yet) worrying about how such perturbations can arise from the environment. However, without an explanation of the origin of the perturbation which includes a dependence on the system state, such an approach has little explanatory power. The Born rule has to be put in by hand by fine-tuning the statistics of the perturbations, and we are left with the feeling that we are missing an essential piece in the model giving it the appearance of a conspiracy theory, rather than anything close to an explanation.

To summarise, whether or not quantum physics (with some assistance of classical physics) can provide an explanation for the occurrence of outcomes in measurements is not yet clear. Arguments which demonstrate the impossibility of outcomes depend on both mathematical and physical idealisations. Physically important conclusions should not hinge on such idealisations. However, when trying to exploit the possibilities when said idealisations are removed, it becomes clear that our understanding of the problem is incomplete at best. The Born rule provides the challenge that we need to meet in order to bridge the conceptual gap.

Like the proverbial bad penny, the Born rule turns up in the next section in very much the same way as it does here. Once again it presents a challenge which needs to be met in order to give physical content to a mathematical view on the problem of outcomes.

\subsection{Problem of Outcomes II} \label{sub:poo1}

Next, we consider a different version of the problem of outcomes. This version initially uses an (exact) measurement basis. The effect of removing this idealisation is subsequently considered. The obtained version of the problem of outcomes could not be used to dismiss the flea model entirely. Yet, this version of the problem does expose an incompleteness in the flea philosophy, as argued in Subsection~\ref{sub:dyn1}.

Let us be explicit about the assumptions. Throughout this section we shall be working within quantum theory. In particular, a measurement is described as a purely quantum mechanical interaction and, unlike the Copenhagen interpretation, none of the degrees of freedom are described in terms of classical physics. We have no problem with applying the formalism of quantum theory to the whole universe, also unlike the Copenhagen interpretation. We do not assume a collapse postulate. We do assume that the formalism of quantum theory, in particular the notion of a state, can be used to describe a single run of an experiment. In other words, we do not assume the need for an ensemble interpretation of quantum theory. 

As before, we assume that there is a 2-level system $\mathcal{S}$, with associated two dimensional Hilbert space $\mathcal{H}_{\mathcal{S}}$ and an environment $\mathcal{E}$ with associated Hilbert space $\mathcal{H}_{\mathcal{E}}$. The environment includes the measuring apparatus and all other degrees of freedom which we might deem relevant, possibly the whole universe if we want.
The initial state of the measurement at a time $t_{0}$ is assumed to be pure, i.e. a normalised vector (up to an uninteresting phase factor) $\rbracket{\Psi_{0}}$ in the Hilbert space $\mathcal{H}_{\mathcal{S}}\otimes\mathcal{H}_{\mathcal{E}}$. Neither the initial reduced system state nor the reduced environmental state is assumed to be pure. By including all relevant degrees of freedom we can assume that the time-evolution is described by a unitary transformation. Let $U=U(t_{0},t_{f})$ denote the unitary time-evolution operator for the total system. The time $t_{f}-t_{0}$ is taken sufficiently long for the measurement to complete and an outcome to be obtained.

We model the measurement using the following assumption about the existence of a measurement basis; there exists an orthonormal basis $\vert m_{1}\rangle$, $\vert m_{2}\rangle$ of $\mathcal{H}_{\mathcal{S}}$, which we call \emph{pointer states}. These are states that become correlated to environmental states, such that in these environmental states the pointer variable is assigned a definite value. Specifically
\begin{equation} \label{eq:pointstat}
U\rbracket{m_{i}}\otimes\rbracket E=\rbracket{m_{i}}\otimes\left[U(m_{i})\rbracket E\right],
\end{equation}
for both $i=1,2$ and any environmental state $\rbracket E$. With respect to the measurement basis, the unitary operator $U$ is represented by a diagonal $2\times2$ matrix of  operators acting on $\mathcal{H}_{\mathcal{E}}$. The operators $U(m_{i})$ on the diagonal are unitary operators on $\mathcal{H}_{\mathcal{E}}$ because $U$ is unitary and diagonal. In particular, (\ref{eq:pointstat}) implies that the unitary operation $U$, when traced over the environment $\mathrm{Tr}_{\mathcal{E}}\left[U\right]$, is diagonal in the pointer basis.

Using the Schmidt decomposition, the initial state can be expressed as
\[
\rbracket{\Psi_{0}}=\sum_{i=1}^{2}c_{i}\rbracket{S_{i}}\rbracket{E_{i}},
\]
where $\rbracket{S_{1}}$ and $\rbracket{S_{2}}$ are orthonormal
basis states for the Hilbert space of the system $\mathcal{S}$, $\rbracket{E_{1}}$ and $\rbracket{E_{2}}$
are orthonormal vectors of the Hilbert space of the environment $\mathcal{E}$, and $c_{i}$
are non-negative real numbers. Consider the action of $U$ on this state, using the pointer basis:
\begin{eqnarray*}
U\rbracket{\Psi_{0}} & = & U\sum_{i=1}^{2}c_{i}\rbracket{S_{i}}\rbracket{E_{i}}\\
 & = & U\sum_{i=1}^{2}c_{i}\sum_{j=1}^{2}\left\langle m_{j}|S_{i}\right\rangle \rbracket{m_{j}}\rbracket{E_{i}}\\
 & = & \sum_{j=1}^{2}\rbracket{m_{j}}\left[\sum_{i=1}^{2}c_{i}\left\langle m_{j}|S_{i}\right\rangle U(m_{j})\rbracket{E_{i}}\right]\\
 & = & \sum_{j=1}^{2}\rbracket{m_{j}}\rbracket{E_{j}^{m}},
\end{eqnarray*}
where in the second line we expanded the system states in terms of
the pointer basis and in the last line we defined the new environment
states $\rbracket{E_{j}^{m}}$. Thus, the density matrix of the system takes the form 
\begin{eqnarray*}
\rho_{S} & = & \left(\begin{array}{c}
\lbracket{m_{1}}\\
\lbracket{m_{2}}
\end{array}\right)^{\dagger}\left(\begin{array}{cc}
\left\langle E_{1}^{m}|E_{1}^{m}\right\rangle  & \left\langle E_{2}^{m}|E_{1}^{m}\right\rangle \\
\left\langle E_{1}^{m}|E_{2}^{m}\right\rangle  & \left\langle E_{2}^{m}|E_{2}^{m}\right\rangle 
\end{array}\right)\left(\begin{array}{c}
\lbracket{m_{1}}\\
\lbracket{m_{2}}
\end{array}\right).
\end{eqnarray*}
The following identity is important, as it implies that the terms $\left\langle E_{i}^{m}|E_{i}^{m}\right\rangle $,
$i\in\{1,2\}$, on the diagonal are \emph{time-independent}. 
\begin{equation} \label{eq:emiemj}
\left\langle E_{i}^{m}|E_{j}^{m}\right\rangle =\sum_{k,l=1}^{2}c_{k}^{*}c_{l}\left\langle S_{k}|m_{i}\right\rangle \left\langle m_{j}|S_{l}\right\rangle \lbracket{E_{k}}U^{\dagger}(m_{i})U(m_{j})\rbracket{E_{l}}.
\end{equation}
Therefore, only the off-diagonal components have a time-dependence (decoherence), since the $U(m_{i})$ are unitary operators.

What does it mean, then, for a measurement to have an outcome? For the moment we adopt the eigenstate-eigenvalue link (e-e link for short): The measurement has outcome $m_{i}$ if and only if the reduced state $\rho_{\mathcal{S}}=Tr_{\mathcal{E}}(\rho(t_{f}))$, of the system, equals the eigenstate $\vert m_{i}\rangle$. In the next section we argue, both from Landsman's asymptotic Bohrification idea and from quantum theory itself, that we need a more general concept of outcome than the one provided by the e-e link. However, in that same section we also address the conceptual problems which arise when we adopt this generalisation.

Adopting the e-e link, the measurement has an outcome only if, relative to the pointer basis, the diagonal components of the reduced density matrix are $(1,0)$ or $(0,1)$, and the off-diagonal components are both 0. Since the diagonal elements of the state are time-independent, we see that in general no single outcome occurs with respect to the basis. This may not be a problem in itself, since we started with an arbitrary state $\vert\Psi_{0}\rangle$, and not every state needs to be a suitable initial state for a measurement. 

The problem is, of course, that, as axiomatised by (\ref{eq:pointstat}), the diagonal elements of the reduced state do not change over time. If at time $t_{f}$ we can assign an outcome $m_{i}$ to the state $U\vert\Psi_{0}\rangle$, then at time $t_{0}$ the reduced state of the system was already equal to $\vert m_{i}\rangle$. The only change that can possibly occur is the disappearance of the off-diagonal terms. As is well known from environmental-induced decoherence, the off-diagonal terms do tend to disappear, but only approximately and temporarily. In short, this model of measurement does not allow for outcomes except for uninteresting trivial cases.

The impossibility of outcomes hinges on the e-e link and the existence of a pointer basis. Is the pointer basis, as previously defined, justified? Although correlating eigenvalues of an operator of $\mathcal{S}$ to macroscopically distinguishable environmental states lies at the heart of a measurement model, assuming that this correlation is completely time-independent is an idealisation. To compensate for the possibility that the pointer basis has a weak time dependence (weak in the sense that the probabilities aren't affected much), we could weaken the condition that $U$ be diagonal to the condition that it is almost diagonal in the sense that
\begin{equation*}
U = \left(
\begin{array}{cc}
U_{11} & U_{12} \\
U_{21} & U_{22}
\end{array} \right),
\end{equation*}
 where we assume that there is a positive number $\epsilon\ll1$ such that the operators $U_{12}$ and $U_{21}$ satisfy $max\{\Vert U_{12}\Vert,\Vert U_{21}\Vert\}\leq\epsilon$ with respect to the operator norm. Note that the operators $U_{11}$ and $U_{22}$ are no longer unitary in that case. On account of being an almost diagonal unitary operator we assume that for $i\in\{1,2\}$, $\Vert U_{ii}^{\dagger}U_{ii}-I\Vert<\epsilon$. Also note that
\begin{equation*}
U\left(\vert m_{1}\rangle\otimes\vert\Phi\rangle\right)=\vert m_{1}\rangle\otimes\left(U_{11}\vert\Phi\rangle\right)+\vert m_{2}\rangle\otimes\left(U_{21}\vert\Phi\rangle\right),
\end{equation*} 
\begin{equation*}
U\left(\vert m_{2}\rangle\otimes\vert\Phi\rangle\right)=\vert m_{1}\rangle\otimes\left(U_{12}\vert\Phi\rangle\right)+\vert m_{2}\rangle\otimes\left(U_{22})\vert\Phi\rangle\right).
\end{equation*} 
Even if the initial state of the system is the eigenstate $\vert m_{i}\rangle$, the post-measurement state is no longer an eigenstate of the observable. If we follow the e-e link to the letter, the observable does not have a value at the end of the measurement. However, the final state is close to an eigenstate in a precise mathematical sense. So it seems tempting to assign the value $m_{i}$ to this state anyway. We discuss the problem of justifying this temptation in subsection \ref{sub:epsi}, and ignore it at the moment. Instead, we ask: What is the effect of using an approximately diagonal $U$ on the problem of outcomes? Equation (\ref{eq:emiemj}) is replaced by a more complicated version which, for $\langle E^{m}_{1}\vert E_{1}^{m}\rangle$, reads as 
\begin{equation*}
\langle E^{m}_{1}\vert E_{1}^{m}\rangle=\sum_{i,j,k,l=1}^{2}c_{i}^{\ast}c_{k}\langle S_{i}\vert m_{j}\rangle\langle m_{l}\vert S_{k}\rangle\langle E_{i}\vert U^{\dagger}_{j1}U_{l1}\vert E_{k}\rangle.
\end{equation*}
In none of the 16 summands the $U$-dependence drops out. Consider for example the summand
\begin{equation*}
S=c^{\ast}_{1}c_{2}\langle S_{1}\vert m_{1}\rangle\langle m_{2}\vert S_{2}\rangle\langle E_{1}\vert U^{\dagger}_{11}U_{21}\vert E_{2}\rangle.
\end{equation*}
By assumption $\Vert U_{21}\vert E_{2}\rangle\Vert<\epsilon$, and 
\begin{equation*}
\vert\Vert U_{11}\vert E_{1}\rangle\Vert^{2}-1\vert=\vert\langle E_{1}\vert(U^{\dagger}_{11}U_{11}-I)\vert E_{1}\rangle\vert<\epsilon.
\end{equation*}
Using the Cauchy-Schwarz inequality we estimate the summand as
\begin{equation*}
\vert S\vert\leq\vert\langle E_{1}\vert U^{\dagger}_{11}U_{21}\vert E_{2}\rangle\vert<\epsilon\cdot\sqrt{1+\epsilon}.
\end{equation*}
For the other 15 summands we can produce estimates in a similar fashion. Writing the initial state as
\begin{equation*}
\vert\Psi_{0}\rangle=\vert m_{1}\rangle\vert E^{m}_{1}(0)\rangle+\vert m_{2}\rangle\vert E^{m}_{2}(0)\rangle,
\end{equation*}
for suitable states $\vert E^{m}_{1}(0)\rangle$, we find
\begin{equation*}
\vert\langle E^{m}_{1}(0)\vert E^{m}_{1}(0)\rangle\vert=\sum_{i=1}^{2}\vert c_{i}\vert^{2}\cdot\vert\langle S_{i}\vert m_{1}\rangle\vert^{2}.
\end{equation*}
From the previous considerations we can obtain the inequality
\begin{equation*}
\vert\langle E^{m}_{1}\vert E^{m}_{1}\rangle-\langle E^{m}_{1}(0)\vert E^{m}_{1}(0)\rangle\vert<\epsilon\cdot\left(4+4\epsilon+8\sqrt{1+\epsilon}\right)<24\epsilon.
\end{equation*}
Recall that $\epsilon\ll1$. Therefore, the Born probabilities associated to the value $m_{1}$ of the observable barely change from the initial to the post-measurement state. For the value $m_{2}$ we can arrive at the same conclusion by analogous means. The time dependence, provided by the non-diagonality of the time-evolution operator is too weak to deliver any non-trivial collapse onto an (approximate) eigenstate for the observable. \\

The previous example does not imply that the problem of outcomes is unsolvable within quantum theory. That would amount to proof by lack of imagination. We should ask, however, whether this argument applies to the flea model, an approach which aims to achieve a collapse by slightly changing the hamiltonian operator over time. The fact that neither the system $\mathcal{S}$ nor the environment $\mathcal{E}$ is made explicit in the model makes this question hard to answer. Even so, in Subsection~\ref{sub:dyn1} we conclude that the flea model needs additional restrictions with respect to initial states in order to avoid the problem of outcomes of the current subsection.

\subsection{The problem of defining outcomes} \label{sub:uitkom}

In this subsection we consider the following issue: Using \emph{pure} states in the formulation of the problem of outcomes contradicts a purely quantum mechanical treatment of this problem. This can be seen as a motivation for using an \emph{approximately} diagonal unitary operator in the previous subsection. More importantly, this issue obstructs our understanding of post-measurement states as outcomes.  

In the discussion of the problem of outcomes we have used that pure states, or density matrices with some zeros on their diagonal, offer an interpretation of certainty on the level of a single system. An interpretation that, for example, a density matrix with a non-zero diagonal does not have. If we can assign a pure state $\vert\psi\rangle$ to the system $\mathcal{S}$, then it is often convenient to think of this state as a catalogue of all the  properties which the system has. Any proposition about the system corresponds to a projection operator $\hat{P}$, and the system has the property corresponding to the proposition iff $\vert\psi\rangle$ lies in the closed subspace corresponding to $\hat{P}$, i.e. $\langle\psi\vert\hat{P}\vert\psi\rangle=1$. However, if the rules of quantum theory are taken seriously, the system state $\rho_{\mathcal{S}}$ is never expected to be pure, and we have already thrown out some information about the environment by restricting to the system. In fact, pure states can only be used by ignoring the ubiquitous entanglement implied by quantum theory and by applying convenient classical idealisations to introduce certainties in a model. We claim that since a pure state is an idealisation, which is used only to sneak in classical certainties into the quantum formalism without proper argument, the use of pure states in any fundamental discussion is unjustified (following Earman's principle, see Section~\ref{sub:bohring}). For an example of often used but unjustified idealisations, which are against the strict formalism of quantum theory, to arrive at pure states or density matrices with some zeros, see Appendix~\ref{sec:nes}. Furthermore, we claim that if the rules of quantum theory are followed to the letter, also density matrices with some zeros on the diagonal are impossible. Simply put, this follows from the fact that all quantum processes that can contribute, will contribute, unless restricted by a conservation law. All these contributions must be taken into account, however insignificant or small they might appear, when taken at face value. Left with only density matrices with non-zero diagonal, the interpretation of certainties in quantum theory completely disappears. In particular, if for a given observable, $\hat{P}$ corresponds to the proposition that a measurement of said observable yields a certain value of the pointer variable, then we expect that $1>Tr(\rho_{\mathcal{S}}\hat{P})>0$, regardless of the value of the pointer variable (as long as $\hat{P}\neq0$). This makes it hard, if not impossible, to think of the system state as a catalogue of properties of a single system, simply because no non-trivial properties satisfy the condition $Tr(\rho_{\mathcal{S}}\hat{P})=1$. 

In classical physics, a state describes the way things are. The view of a quantum state as a collection of properties of the system seems the closest that we can get to this classical understanding of states. Yet, the rules of quantum theory force us to abandon this perspective. But if we choose to do this, how should we think of states in models of quantum measurement instead? Contrary to most of the work in foundations of quantum mechanics, work on models of quantum measurement \cite{abn,hasp} tend to favour an ensemble interpretation of quantum mechanics. Using states only at the level of ensembles is undesirable in that it may entail that the problem of outcomes becomes unsolvable within the confines of quantum theory. Even so, problems such as our inability to assign properties to a system and the problem of independence, which is discussed later, may turn out to be artefacts due to a wrong state ontology, and may be alleviated within an ensemble interpretation. 

The connection between actual system preparations in the lab and the formalism of quantum mechanics is a complex one, and the question whether we \emph{need} an ensemble interpretation is a subject for another paper entirely. We mention it here, as the mathematical formulation of the problem of outcomes depends on the state ontology. In any case, any attempt to solve the problem of outcomes from within quantum theory should first rigorously define what it means for a single system to have a property within the theory, i.e., define a truth indicator, and take care not to resort to idealisations which contradict quantum theory itself. 

\subsection{The problem with Epsilonics} \label{sub:epsi}

Here we discuss a problem that is encountered in almost all attempts at solving the measurement problem and we will run into it in the next section when discussing the flea proposal. It deals with an often used way of arguing that some states are close to pure states, or density matrices with some zeros on their diagonal, in order to arrive at statements of certainty (outcomes). As we hinted at in the previous subsection, outcomes and measurements are clear in the classical domain, however, we should seek their explanation in quantum theory. Within quantum physics, at best we get $\epsilon$-approximate outcomes, in the form of states, where a high probability of $1-\epsilon$ can be assigned to a single outcome, and the other possible outcomes contribute a small albeit non-zero probability no larger than $\epsilon$. The smallness in the difference between $\epsilon$-approximate outcomes and outcomes is made precise mathematically. However, we also need to understand this smallness physically, before we can conceptually connect $\epsilon$-approximate outcomes to outcomes. It is only by the nature of the questions which we are asking that this is in fact a problem. We are trying to understand outcomes, and, hopefully, even the Born rule. It is therefore circular to a priori assume the Born rule. Yet, without the Born rule, is there any argument that we can invoke to motivate that the $\epsilon$-small deviations of outcomes are insignificant?\\

This problem of epsilonics should remind us of a similar problem faced in attempts to derive the Born rule in the many worlds interpretation. As with the flea proposal, the formalism of quantum physics takes a central position in resolving the problem of outcomes, within this approach. In fact, it is often claimed that the many worlds interpretation, or everettian quantum mechanics as it is more often called by its practitioners, is the interpretation which is naturally implied by the mathematical formalism.  Also, like the flea proposal, the claim is that there is no actual problem of outcomes, in the sense of a contradiction following from the existence of outcomes and the predictions from the quantum formalism. In everettian quantum mechanics, the \emph{apparent} conflict arises from a misguided understanding of the formalism, of superpositions in particular, in lieu of mathematical idealisations being the culprit in the flea proposal. In everettian quantum mechanics there is no measurement problem, and the Born rule can allegedly be derived. The derivation of the Born rule has faced much criticism, from various directions, but here we concentrate only on one line of arguments which is relevant to the problem of epsilonics. The problem is that through its reliance on decoherence, the derivation becomes circular, as first noted by Zurek \cite{zurek1, zurek2}, and later by Baker \cite{baker}, and Kent \cite{kent}. We concentrate on the recent criticism by Dawid and Th\'ebault \cite{tida}. These authors argue that the use of decoherence in the derivation of the Born rule is not merely circular, but even leads to a conflict between the fundamentals precepts with regard to the different roles that probability plays in the approach. This allegedly renders the whole everettian approach inviable. Since this inviability relies on the subjective nature of the decision theoretic version of probability specific to everettian quantum mechanics, it is not important for our current purposes. Our main reason of using \cite{tida} rather than \cite{zurek1,zurek2,baker,kent} is that their presentation fits well with the problem of epsilonics for asymptotic Bohrification.

In order to remove macroscopic superpositions, and to invoke classical decision theory, decoherence is needed in everettian quantum mechanics, for only under the assumption of decoherence does a \emph{quasi}-classical branching structure emerge in the formalism of quantum physics, where only unitary dynamics is allowed. Unfortunately, environment-induced decoherence does not completely remove the interference terms. These off-diagonal terms, with respect to the pointer basis, only become small for all practical purposes, and only for a finite time. Consequently, the discrete branching ontology of everettian quantum mechanics is not realised by decoherence, but only approximated. The problem is to bridge the gap between the quasi-classical branching delivered by decoherence, and the exact branching ontology without smuggling in new concepts, such as the Born rule. For it is by the Born rule that we can justify that the small amplitudes appearing in the off-diagonal components, can be safely ignored. Smallness of the numbers by itself does not provide any justification, especially if we are not clear on what these numbers mean.

Now we find us at the circularity argument of Zurek. Everettian quantum mechanics seeks to derive the Born rule and it needs decoherence for its ontology. Decoherence only provides an effective or approximate account of the discrete branching. To pass from this effective description to the setting where classical decision theory can be used, we need to assume the Born rule. Proponents of everettian quantum mechanics have, of course, been aware of such circularity arguments. Wallace \cite{wall}, might dismiss the circularity argument by denying that the last step is needed. According to Wallace the discrete branching structure of everettian quantum mechanics is understood as a \emph{robust} yet \emph{emergent} feature of reality. As is typical for discussions surrounding questions of emergence and reduction, it is hard to understand what this robustness means exactly. 

Dawid and Th\'ebault \cite{tida} argue along analogous lines with regards of robustness, concentrating on \emph{empirically grounded} robustness, as the only kind that holds explanatory power. In their words:
\begin{quote}
The first crucial distinction that can be made is between a notion of robustness that is empirically grounded and one that is not. By this we mean some qualification such that whether a structure within the formalism of a theory is taken to be robust is dependent upon some interpretational connection between that structure and empirical phenomenology.
\end{quote}

For everettian quantum mechanics, the Born rule provides the empirical grounding needed before one can invoke robustness as an argument carrying any conceptual weight. \\

This concludes our small excursion through everettian quantum mechanics. We close this subsection with a brief discussion of the tail problem in GRW models for dynamical collapse~\cite{bagi}. Recall that these models supplement unitary quantum theory by a non-linear and stochastic spontaneous collapse process. First of all, a warning is in order. We initially consider the dynamical collapse models from the perspective of standard quantum theory (be it with an adapted mathematical framework), rather than from the matter density ontology introduced in~\cite{ggb}. This difference in ontology is relevant to the tails problem, and therefore to the problem of outcomes. 

Suppose we are given an initial state $\vert\Psi\rangle=\sum_{i=1}^{n}a_{i}\vert x_{i}\rangle$, with non-zero coefficients $a_{i}$, and the states $\vert x_{i}\rangle$ correspond to different positions where a certain particle can be found.  In the idealised theory, such a state should collapse to a state where all but one of the coefficients are reduced to zero. But this does not happen in the GRW models, and it is even considered unphysical to desire this. As formulated in~\cite{macq}:

\begin{quote}
This is due to position/momentum incompatibility. The more confined the position wave-function, the more spread out the momentum wave-function. The more spread out the momentum wave-function, the more equiprobable all possible states of momentum become. The relationship between energy and momentum then yields drastic post-collapse violations of energy conservation: ones that we know by experiment do not occur. So GRW formulated the collapse function as a gaussian.
\end{quote}

Collapsed states in GRW are gaussian and therefore the associated wavefunctions are non-zero throughout the whole space. From the point of view of standard quantum theory, this once again raises the question of how such states can be understood as outcomes. We are facing the same problem as before, for asymptotic Bohrification. For Ghirardi, Grassi and Benatti, the tails do not form a problem, as these are defined away in the matter density formulation of the theory~\cite{ggb}. The tails of a collapsed state correspond to low-density matter regions of the wavefunction, which are deemed not relevant because they are inaccessible. But what does it mean for a matter disribution to be accessible or inaccessible? If inaccessible means that any observer is unable to measure it, then we share the worries expressed in \cite{tumu, macq} that we should not rely on observers before the tail problem is resolved. 

Since our point was to demonstrate the ubiquitous use of approximations hinders us in our understanding of post-measurement states as outcomes, at least without the Born rule, we leave it at this. In the next section we will encounter many of the issues discussed in this section for the particular case of the flea proposal.

\section{Bohrification and Approximate Outcomes} \label{sec:bohr}

In Subsection~\ref{sub:bohring} we introduce asymptotic Bohrification, wherein lies the formulation of the problem of outcomes, which the flea model aims to solve. This formulation of the problem of outcomes requires us to rely on an approximate version of outcome states, without providing physical grounding of how to understand such states. As was discussed in Subsection~\ref{sub:epsi}, such approximate outcomes states are typical of approaches to the measurement problem, which take the formalism of quantum mechanics seriously. In fact, as can be seen from the example in Appendix~\ref{sec:nes}, such approximate outcomes states, tricky as they may be conceptually, are needed for any quantum mechanical treatment of the problem of outcomes which does not use classical approximations. Finally, in Subsection~\ref{sub:dis3} we briefly review the problems (encountered up to this point) in solving the problem of outcomes within the formalism of quantum theory.

\subsection{Asymptotic Bohrification} \label{sub:bohring}

Consider the following incarnation of the problem of outcomes, adapted from~\cite{maudlin}. According to Maudlin, the problem of outcomes is the incompatibility of the following three assumptions:
\begin{enumerate}
\item Quantum mechanical pure states are complete in the sense that they specify all physical properties of a system.
\item Time-evolution of the states is described by a linear unitary operator.
\item Measurements always (or at least usually) have single outcomes, i.e. at the end of the measurement, the measuring device indicates a definite physical state.
\end{enumerate}
In the previous section we concluded that, if we remove certain mathematical idealisations, then it is not clear whether these combined assumptions yield a contradiction. Regardless, there is still a challenge to be met in combining these assumptions. The various approaches to the problem of outcomes differ in which of the above three assumptions are challenged. In hidden variable models the first assumption is denied. For dynamical collapse theories such as the GRW models, the second assumption is rejected. For the many-worlds interpretations the third assumption gets a new perspective. But how would the practitioner of the Copenhagen interpretation consider this problem? He (or she) would most likely frown at the first assumption. Properties are classical concepts and have no place in quantum theory. For now, let us ignore this issue. At this point, the Copenhagenist may claim that there is no problem of outcomes, in the guise of a contradiction. Indeed. assumptions (1) and (2) are about the quantum mechanical formalism whereas assumption (3) deals with the purely classical notions of outcomes and measurements. If we are to connect these assumptions, then we need to use classical approximations at some point of the description. It is exactly because of the need of these approximations that the irreducible probabilities of quantum theory arise. There is nothing mysterious about these probabilities, in the sense that in a purely quantum mechanical description probabilities need not arise. However, aside from putting us in the awkward position that a theory which supposedly generalises classical physics, also needs classical physics for its formulation, the Copenhagen move teaches us little, if anything, about measurements in quantum theory. And so we ask; why would we \emph{need} classical approximations in the first place? Consider Bohr's doctrine of classical concepts:
\begin{quote}
However far the phenomena transcend the scope of classical physical explanation, the account of all evidence must be expressed in classical terms. (…) The argument is simply that by the word experiment we refer to a situation where we can tell others what we have done and what we have learned and that, therefore, the account of the experimental arrangements and of the results of the observations must be expressed in unambiguous language with suitable application of the terminology of classical physics.
\end{quote} 
Although this strikes us as common sense, there are different ways in which we can implement this philosophy with respect to the problem of outcomes. Note that if we accept the need of using the \emph{language} and experience from classical physics, this does not automatically entail that we need to replace approximations from the formalism of classical mechanics in studying the problem of outcomes. Just to be clear, we do not see wisdom in attempting to describe any realistic measurement apparatus (and environment) completely at the level of the standard model, and to use this as a model of measurement. However, we could endorse Bohr's doctrine of classical concepts, without making the jump to the rest of the Copenhagen interpretation, taking the stance that the language of classical physics is essential to the problem of outcomes, but, unlike the Copenhagen interpretation, not \emph{a priori} assuming that the occurrence of an outcome cannot be explained by using a suitable quantum-mechanical model. 

This brings us to the \emph{Bohrification} approach to the problem of outcomes, proposed by Landsman. In this approach, Bohr's doctrine of classical concepts is given the following mathematical interpretation: study non-commutative C*-algebras, such as the algebra of all bounded operators on a Hilbert space, by means of commutative C*-algebras. There are two different ways in which this is done. The first approach, called \emph{exact Bohrification} replaces a non-commutative C*-algebra by its partially ordered set of commutative C*-subalgebras, where the order is given by inclusion. Exact Bohrification is the central theme of the contribution of Landsman and Lindenhovius to this volume. Since this approach has not been applied to the problem of outcomes, we shall not consider it any further. In the second approach, called \emph{asymptotic Bohrification} the commutative and non-commutative C*-algebras, no longer related by an inclusion relation, are glued together in a bundle called a countinuous field of C*-algebras. Rather than discuss this approach in full generality, we first consider the motivating example of the flea approach to the problem of outcomes, as introduced by Landsman and his student Reuvers in~\cite{lare}, and later embedded by Landsman in the asymptotic Bohrification programme in~\cite{landsman}. The terminology which we adopt here was introduced by Landsman in~\cite{landsman3}.

Consider the following simplistic model used to reformulate the problem of outcomes. The relevant Hilbert space is $\mathcal{H}=L^{2}(\mathbb{R})$. The dynamics is generated by the hamiltonian
\begin{equation} \label{equ_hamop}
\hat{H}_{\hbar}=-\frac{\hbar^2}{2m}\frac{d^2}{dx^2}+\frac{\lambda}{8}\left(x^2-a^2\right)^2,
\end{equation}
using a symmetric double well potential, with barrier height $V_{\mathrm{b}}=\lambda a^{4}/8$. The subscript $\hbar$ was added because we consider different values of $\hbar$, and are in particular interested in the limit $\hbar\to0$. In this model, the value of $\hbar$ is used to represent a scale for macroscopicity, where smaller values of $\hbar$ correspond to more macroscopic situations. The reader who is uncomfortable varying a constant of nature may instead consider the limit $\lambda\to\infty$ where the potential energy becomes steeper, or the limit $m\to\infty$.

\begin{figure} \label{fig:flappybird}
\begin{centering}
\includegraphics[scale=0.4]{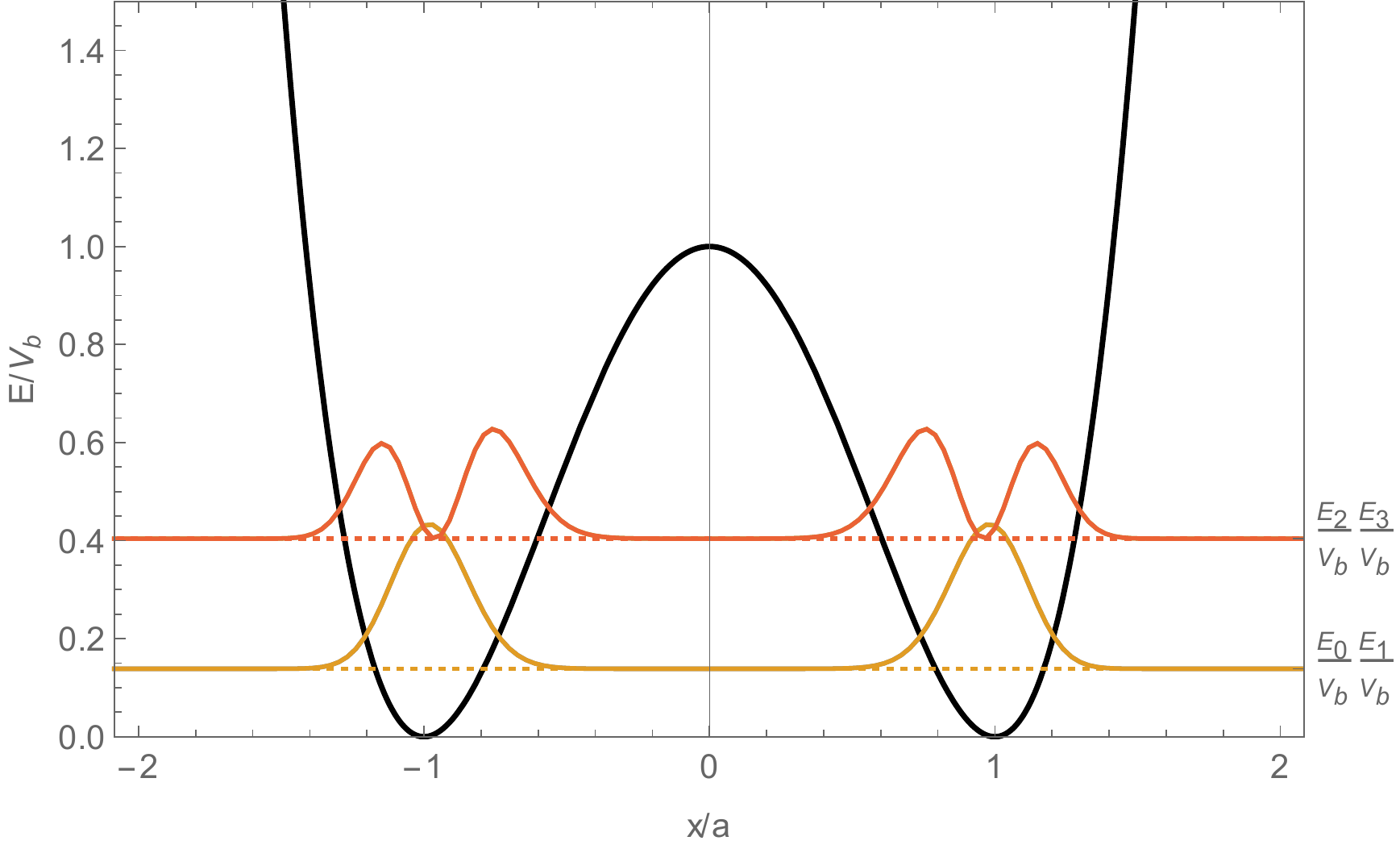} 
\par\end{centering}

\protect\protect\caption{The first four eigenfunctions as $\left|\psi(x)\right|^{2}$ for the
double well potential with $l\ll a$. The first and second, and the third and fourth, wavefunctions overlap due to the symmetry.}
\end{figure}

The initial state for the model is the (non-degenerate) ground state $\psi_{0}$ for this hamiltonian. The characteristic length of the problem, determining the scale on which the wavefunctions vary, is $l=(\hbar^{2}/mk)^{1/4}$. Here $k=\lambda a^{2}$ is the spring constant as determined by the quadratic approximation at the well minima $x=\pm a$. 
Consider the case with fixed $\lambda,a$ but variable $\xi=\sqrt{\hbar^{2}/m}$. We are interested in the setting where $l\ll a$, i.e., large mass or small $\hbar$, where the lowest energy eigenstates are localised within the two wells, see Figure~\ref{fig:flappybird}. 

The ground state $\psi_{0}$ is to represent the state of a pointer or some other macroscopic variable at some point of time during the measurement interaction. Supposedly, there is a system $\mathcal{S}$ which was initially in a Schr\"odinger cat state. Through its interaction with the measurement apparatus, this Schr\"odinger cat state was passed on to the pointer. The two wells in the potential correspond to two macroscopically distinct values that the pointer variable may take. The $x$-variable need not correspond to a physical distance; we only identify the two wells with two pointer values. The system $\mathcal{S}$ itself is not visible in this simple model. Neither does the model include an environment $\mathcal{E}$, which contains the uncontrollable degrees of freedom of the apparatus and any other degrees of freedom we may consider to be of interest. The environment plays an important conceptual role in the flea approach to the problem of outcomes, but the environmental degrees of freedom do not appear in the model themselves. 

It may strike the reader as odd to use a bound state, the textbook example of a stable state, as an initial state of the model. Keep in mind that the initial state of the model is not the initial state of the measurement. The initial apparatus state may very well have been metastable, as is typical for models of quantum measurement. In addition, we shall see that the flea approach uses a time-dependent hamiltonian, so the initial state does not remain a bound state. Regardless, we may still be sceptical whether the setting of a bound state sporting a macroscopic superposition actually occurs during \emph{any} measurement. We postpone further discussion with regards to the justification of the model to the next section.

At this point we can discuss the reformulation of the measurement problem. The initial state $\psi_{0}(x)$ has a $\mathbb{Z}_{2}$-symmetry, through reflection in the $y$-axis, a symmetry inherited from the invariance of $\hat{H}_{\hbar}$ under $x\mapsto -x$. Without any further additions or modifications (such as the flea) the wave function does not change over time and retains this symmetry, even when it becomes the post-measurement state. This happens for all non-zero values of $\hbar$. In the limit $\hbar\to0$, the post-measurement state $\psi_{0}(x)$ converges, in a sense we make precise mathematically in a moment, to the following classical mixed state
\begin{equation} \label{equ_mixmaster}
\rho_{0}^{(0)}=\frac{1}{2}\left(\rho_{0}^{+}+\rho_{0}^{-}\right),
\end{equation}
where the two phase space points
\begin{equation*}
\rho_{0}^{\pm}=\left(p=0,q=\pm a\right)
\end{equation*}
are the two classical ground states of the (classical) hamiltonian
\begin{equation}
h_{0}(p,q)=\frac{p^{2}}{2m}+\frac{\lambda}{8}\left(x^2-a^2\right)^2.
\end{equation}

From the point of view of asymptotic Bohrification, the problem of outcomes is that the quantum-mechanical post-measurement pure state converges to a classical mixed state. Or, alternatively stated, it is not the case that for sufficiently large mass $m$ or small $\hbar$, the post-measurement state approximates a classical pure state. The problem of outcomes can be seen as the incompatibility of the following three assumptions:
\begin{enumerate}
\item Measurements and their outcomes are notions from classical physics.
\item In many cases of interest, the transition from quantum physics to classical physics can be described by a limits such as $\hbar\to0$ (or, to be briefly considered at the end of the following section,  $N\to\infty$, where $N$ is the number of degrees of freedom in the model).
\item Whenever such a limit is applicable, any physical effect in classical physics must be foreshadowed in quantum physics.
\end{enumerate}

Let us consider the third assumption, which is vital. Even though the notion of outcome only has a meaning in the limiting classical theory, the classical limit itself is an idealisation and any phenomenon cannot be counted as genuinly physical if it only appears in this idealised limitting theory. The philosophy adopted in asymptotic Bohrification, telling us that outcomes should have approximate quantum-mechanical counterparts, is partly captured by \emph{Earman's principle} \cite{oormans}:

\begin{quote}
While idealizations are useful and, perhaps, even essential to progress in physics,
a sound principle of interpretation would seem to be that no effect can be counted
as a genuine physical effect if it disappears when the idealizations are removed.
\end{quote}

The rest is captured by \emph{Butterfield's Principle} \cite{boterveld}:

\begin{quote}
there is a weaker, yet still vivid, novel and robust behaviour that occurs before we get to
the limit, i.e. for finite $N$. And it is this weaker behaviour which is physically real.
\end{quote}

The problem of outcomes, as it arises in the simple double well model, amounts to the problem that for small non-zero values of $\hbar$, the post-measurement state does not \emph{approximate} one of the classical pure states $\left(p=0,q=\pm a\right)$, which we identify as outcomes in this model. Mathematically, the term \emph{approximate} has a precise meaning in asymptotic Bohrification, expressed in the language of continuous fields of C*-algebras. The definition can be found in Appendix~\ref{sec:convergence}.

It is crucial to understand the way in which the post-measurement state should approximate an outcome. By assumption, the very notion of outcome is a classical one. Yet, in following Butterfield's principle we should concentrate on quantum mechanics for small values of $\hbar$ (or in another suitable limit). It is the notion of convergence that tells us in which way the quantum mechanical states should be close to classical outcomes. Therefore, in addition to a precise mathematical formulation, we need a physical grounding for convergence of states. The following quote, taken from~\cite{landsman2}, should help in providing insight in the physical grounding of convergence of states, as it explains the way that the doctrine of classical concepts is understood in the operator algebraic setting of asymptotic Bohrification. The quantisation map $Q_{\hbar}$ used in this quote is defined in Appendix~\ref{sec:convergence}.

\begin{quote}
The map $\mathcal{Q}_{\hbar}$ is the quantization map at value $\hbar$ of Planck's constant; we feel it is the most precise formulation of Heisenberg's original \emph{Umdeutung} of classical observables known to date. It has the same interpretation as the heuristic symbol $\mathcal{Q}_{\hbar}$ used so far: the operator $\mathcal{Q}_{\hbar}(f)$ is the quantum-mechanical observable whose classical counterpart is $f$.
\end{quote}

Classical physics enters the problem of outcomes through deformation quantization rather than through classical approximations, as used in the Copenhagen interpretation. The umdeutung of the quotation is central to asymptotic Bohrification. Because of this, one might think that when a salesman, selling asymptotic Bohrification, is at the door, rather than Bohr, Heisenberg is the one who knocks.

\begin{figure} \label{fig:fleabird}
\begin{centering}
\includegraphics[scale=0.4]{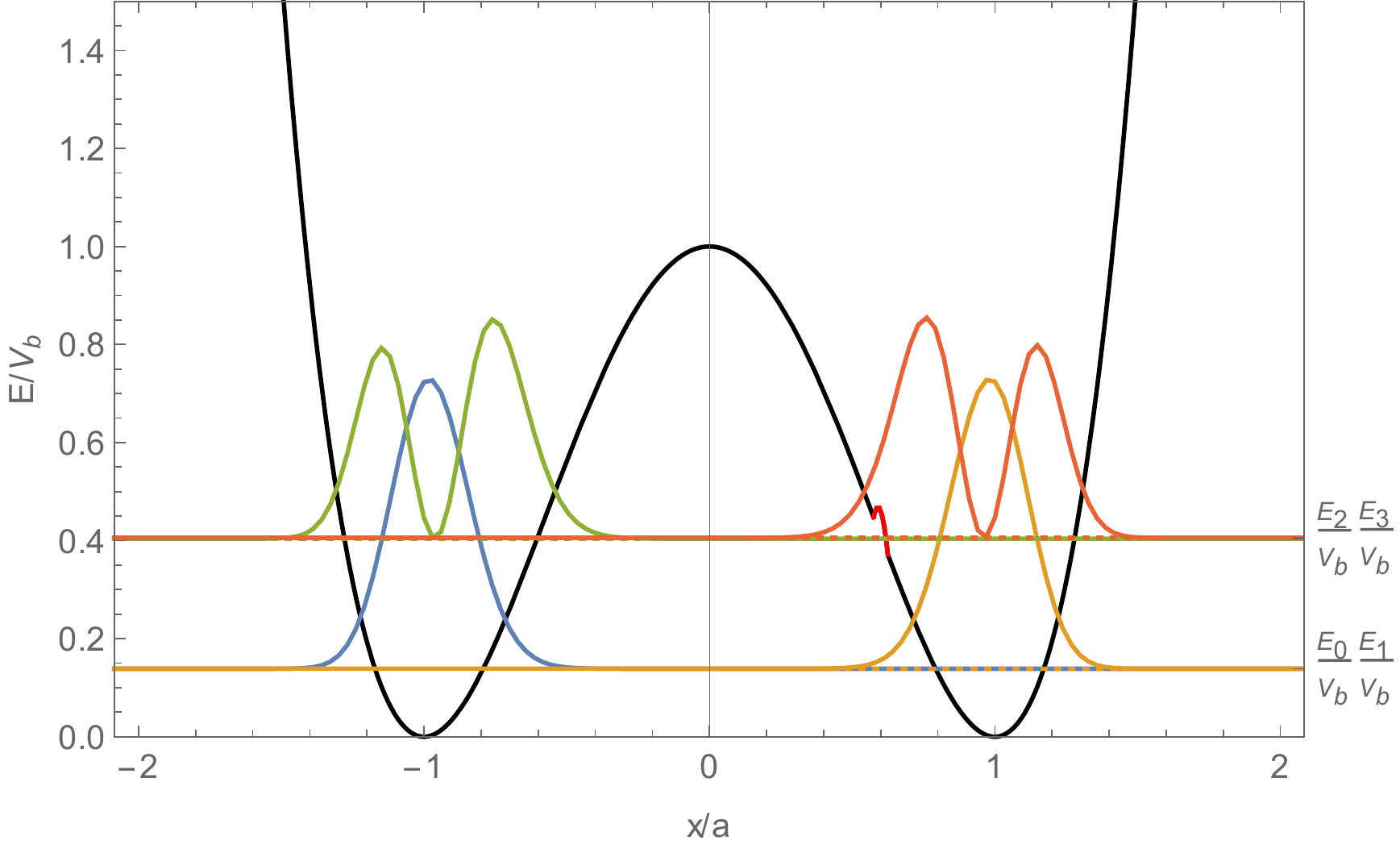} 
\par\end{centering}

\protect\protect\caption{The first four eigenfunctions as $\left|\psi(x)\right|^{2}$ for the
double well potential with flea perturbation (marked red). From left to right, bottom to top, the maxima give the first (blue), second (yellow), third (green) and fourth (red) wavefunctions.}
\end{figure}

Although there is some physical underpinning for the quantum mechanical approximate outcomes, we argue that this understanding is incomplete, at least for solving the problem of outcomes. To make the discussion more concrete, briefly consider the flea proposal \cite{lare}, intended to alleviate the problem of outcomes. The time-independent hamiltonian of (\ref{equ_hamop}), which, from now on, we denote as $\hat{H}_{\hbar,0}$ is replaced by a time-dependent one $\hat{H}_{\hbar}(t)=\hat{H}_{\hbar,0}+f(t)\delta V$, where $\delta V$ is a small perturbation of the potential, localised in one of the two wells, and $f(t)$ is some scalar function changing the hamiltonian from $\hat{H}_{\hbar,0}$ initially, to $\hat{H}_{\hbar,0}+\delta V$. The motivation for this move can be found in the work of Jona-Lasinio et.al. \cite{jola} regarding the sensitivity of ground states with respect to small perturbations in the semi-classical setting. Provided that $\hbar$ is taken to be small enough, even for a minute perturbation the ground state of $\hat{H}_{\hbar,0}+\delta V$ is highly concentrated in one of the two wells, as in Figure~\ref{fig:fleabird}. 

The idea is that for a suitable dynamical introduction of the flea, the wave function, initially expressing a Schr\"odinger cat state, evolves to the ground state with the perturbed potential resulting in a post-measurement state which is highly concentrated in one of the two wells. In the next section we explore the feasibility of such an approach. In this section however, we ask a different question. Assuming that the flea does its job, how much closer would that bring us to solving the problem of outcomes? The point is that although the obtained ground state is largely concentrated in a single well, the probability associated to the other well is still non-zero. In writing off this non-zero probability as unimportant, we are in fact saying that the obtained state is sufficiently like a (classical) outcome. However, we should be very careful in what we are assuming when we state that such small probabilities do not matter. We seek to explain how a single measurement run can have an outcome. As argued in the previous section, a satisfactory resolution of this problem should also bring us much closer to an understanding of the Born rule; otherwise we are left with odd conspiracies which challenge the very notion of independence, lying at the heart of the very notion of measurement. 

Previously, we argued that the problem of small probabilities need not be seen as a weakness of asymptotic Bohrification in the sense that any approach to the problem of outcomes, which does not use classical approximations but seeks to work within quantum mechanics, is forced to consider approximate outcomes too, and therefore needs to address the problem of small probabilities.

\subsection{Discussion} \label{sub:dis3}

Before moving further with the flea model, let us consider the main issues up to this point.

\subsubsection{Problem of defining outcomes}

\begin{cha}[Problem of defining outcomes]
In an attempt at solving the measurement problem, the outcome of a measurement must be well-defined and not rely heavily on idealisations. For attempts at solving the measurement problem from within quantum theory, a notion of an outcome must be defined which does not rely on classical idealisations in the model that remove potential sources of entanglement or completely restricts the contributions to certain outcomes. In particular, pure states and density matrices with some zeros on the diagonal must be avoided, or at least extensively scrutinized as they risk contradicting quantum theory itself.
\end{cha}

Even though the definition of an outcome is arguably the most crucial in a discussion of the measurement problem, this problem had not been sufficiently addressed in asymptotic Bohrification. This proposal has the benefit that in a certain limit outcomes are well-defined classically, however, as stated in Subsection~\ref{sub:uitkom}, we need a quantum mechanical definition of an outcome. Adherence to Butterfield's principle puts further emphasis on understanding outcomes quantum mechanically.

Furthermore, for its clear classical limit it assumes that the groundstate is a pure state initially (pre-collapse) and finally (outcome). According to the ubiquitous entanglement of quantum theory and the supposed size of the system the groundstate represents, it seems counter-intuitive that both the initial and final state are not expected to be entangled in some way. Especially if the flea is of quantum mechanical origin and it comes from the environment, one would expect that a full quantum treatment would lead to entanglement between the system the groundstate represents and the environment.
 
We will further discuss the implications of this assumption at the start of Section~\ref{sec:flea}.

\subsubsection{Problem of Epsilonics}
\begin{cha}[Problem of Epsilonics]
In a fully quantum mechanical treatment of a measurement, the post-measurement state is not a state where we can assign a single value of the pointer variable with absolute certainty, but only with a probability of $1-\epsilon$, for some $0<\epsilon\ll1$. The initial state for the measurement is never an eigenstate for the observable (the operator to be correlated with the pointer variable). It has a non-zero probability associated to each value of the observable, and is expected to be entangled with the environment. In order to think of the final state as an outcome, and the initial state as carrying properties of the system, we are required to assume the Born rule.
\end{cha}

The pragmatic physicist can safely ignore this problem and, e.g. trade the mathematically cumbersome initial states where the system is entangled with the environment in for a state such that the reduced system is state is assumed pure. For a well designed preparation method, the difference between the eigenstate and the exact state is negligible as far as the statistics of outcomes of the subsequent measurements are concerned. However, when targeting the specific foundational issue of explaining the occurrence and statistics of outcomes, such a move may very well throw the baby out with the bath water. Of course, if our philosophical underpinning would be that of the Copenhagen Interpretation, then we could make the same approximations as the pragmatist, but now motivated by treating part of the setup as classical. But this approach renders the probabilities irreducible, thereby placing the problem of outcomes beyond the reach of quantum mechanics. 

Prima facie, the epsilonic contributions in the approximate outcome states do not matter, since such contributions vanish in the limit $\hbar\to0$. Robustness in this setting, is what survives in the $\hbar\to0$ limit. This is a mathematical condition, and as noted in the previous section, we need a physical grounding of convergence (or robustness) in order to understand how these approximate outcomes (which are of central interest by Butterfield's principle) are connected to the idealised classical outcomes. In the toy double well example from asymptotic Bohrification, where classical physics is emergent from quantum theory, we see that scaling $\hbar$, using a slightly perturbed potential, correlates with bound states which become heavily peaked in a single well. Again, it is the Born rule that provides the empirical grounding to this mathematical notion of convergence. 

Previously, we emphasised that the problem of epsilonics requires us to a priori assume the Born rule. Alternatively, we can concentrate on the consequence that we cannot think of an initial state as a collection of properties of the system. In this light, the problem of epsilonics can be viewed as an indication that we are asking the wrong questions and are using a wrong mathematical formulation for the problem of outcomes. 

\subsubsection{Problem of Independence}

Why do we consider a priori assuming the Born rule to be an issue? Even if we are to assume the Born rule, to provide the desired physical grounding to e.g. asymptotic Bohrification, solving the problem of outcomes as formulated in this approach seems like a huge stride forwards for the foundations of quantum mechanics. The main reasons that we are so concerned with the Born rule can be found in Section~\ref{sec:mp}.

Consider Subsection~\ref{sub:poo2} where we first considered the problem of outcomes. By dismissing counterfactual reasoning, as well as removing the idealisation of using exactly the same unitary operator and initial environmental states, the contradiction between outcomes of measurements and the formalism of quantum theory disappeared. However, in Subsection~\ref{sub:poo1} it was shown that only slightly varying the unitary operator is insufficient for solving the problem of outcomes. As argued in Subsection~\ref{sub:dyn1} , for the case of the flea, de facto superselection rules for initial states are needed in order to achieve an effective collapse of the wave function. The choice of initial environmental states and the variations in unitary dynamics dictate the outcomes, leading to the following problem.

\begin{cha}[Problem of Independence]
For measurement interactions described within unitary quantum mechanics, the statistics of outcomes is determined by the statistics of the variations of initial states and the unitary time-evolution operator. For a fixed system preparation method, despite variations of the initial state for different runs, the Born probabilities associated to the observable do not vary significantly. These Born probabilities, determined by the reduced system state of any single run, also determine the statistics of outcomes. There is no explanation why the Born probabilities should match the probabilities obtained from the statistics of variations. In particular, if, like in the flea model, the variations in the time-evolution operator are expected to be independent of the system under investigation.
\end{cha}

At best, this problem states that the flea model is incomplete and that the Born rule has to be put in by hand by matching the statistics of variations to the desired Born probabilities. However, when we conclude in Subsection~\ref{sub:genfl} that the flea model is not feasible as a model of dynamical collapse, the problem of independence forms the main obstruction in improving the model. Currently, the Born rule is the only known connection between the system and the flea-like variations of the environment. Here lies our reluctance in a priori assuming the Born rule for the problem of epsilonics. In absence of models where a flea actually emerges from an environment, reproducing the Born rule is the only guide in putting some physical content to the flea variations. 

\section{Flea Model} \label{sec:flea}

\subsection{First Remarks}

We return to the setting of Subsection~\ref{sub:bohring}. The Hilbert space is $\mathcal{H}=L^{2}(\mathbb{R})$, and the initial state is the symmetric ground state of the hamiltonian
\begin{equation}
\hat{H}_{\hbar}=-\frac{\hbar^2}{2m}\frac{d^2}{dx^2}+\frac{\lambda}{8}\left(x^2-a^2\right)^2,
\end{equation}
The two wells of the potential represents two distinct values which the pointer variable of the measurement device can assume, and the symmetric ground state represents a Schr\"odinger cat like state, presumably transferred from the state of some quantum two-level system. The starting point of the flea model is supposed to take place near the end of a measurement interaction. The flea is a small (in sup-norm) asymmetric potential $W(x)$ which is localised near the bottom of one of the two wells of the symmetric double well potential $V(x)$. For a small value of $\hbar$, the ground state of the perturbed hamiltonian $\hat{H}_{\hbar}+W$ is nigh completely localised in a single well. The challenge of the flea model is to find a time dependence for the flea, e.g. a function $t\mapsto f(t)$, in such a way that the initial symmetric wave function evolves to the localised ground state of the perturbed setting, as the hamiltonian evolves as $t\mapsto \hat{H}_{\hbar}+f(t)W$. We consider this challenge of dynamics in Subsection~\ref{sub:dyn2}. In this subsection we ask a different question: How is it possible that in the setting of quantum measurements, where entanglement plays such an important role, the flea model describes the dynamics of the pointer in terms of a single pure state?

The flea model assumes that, from the moment at which the flea is introduced, up to the end of the measurement, the degrees of freedom which are explicitly modelled are described by a pure state. It is tempting to think of the wave-function under investigation as representing the state of the pointer variable, but things cannot be that simple. Certainly, there is a relation between the pointer variable and the wave function since the two wells of the potential correspond to the two possible values of the pointer variable. However, if we think of the flea model as being obtained from a more complete measurement model by tracing out the system and environmental degrees of freedom, then we would expect entanglement between the system and the pointer variable to result in a non-pure state. This is crucial to the flea model since it relies on properties of bound states. First, consider the time evolution for a von Neumann ideal measurement
\begin{equation} \label{eq:fleabash}
\frac{1}{\sqrt{2}}\left(\vert m_{1}\rangle+\vert m_{2}\rangle\right)\otimes\vert r\rangle\mapsto\frac{1}{\sqrt{2}}\left(\vert m_{1}\rangle\otimes\vert  E_{1}\rangle+\vert m_{2}\rangle\otimes\vert E_{2}\rangle\right).
\end{equation}
Here $\vert r\rangle$ is the initial environmental state, and the environmental states $\vert E_{i}\rangle$ are close to orthogonal. When we trace out everything but the pointer, the entanglement with the system causes the end result to be a non-pure state. However, as in Subsection~\ref{sub:poo2}, we cannot accept (\ref{eq:fleabash}) at face value since it was derived using counterfactual reasoning, or assuming an eigenstate-eigenvalue link. Different runs of the measurement, possibly with different system preparations, correspond to different initial environmental states. If the environment includes enough degrees of freedom, then for initial states of the form\footnote{According to Appendix~\ref{sec:nes}, it is more accurate to replace the simple tensor $\vert\phi\rangle\otimes\vert r_{\phi}\rangle$ by a density operator which is close to this pure state in norm. Since this change will not affect the conclusions of this subsection, we shall not further consider it.} $\vert\phi\rangle\otimes\vert r_{\phi}\rangle$, we expect that $\langle r_{\phi} \vert r_{\psi}\rangle\approx0$ whenever $\vert\langle\phi\vert m_{1}\rangle\vert\neq\vert\langle\psi\vert m_{1}\rangle\vert$, because the preparation set-up for the two system states is macroscopically distinct. However, we expect such differences to be largely irrelevant\footnote{Although the problem of independence may make us reconsider this. At this point however, we shall not entertain such thoughts which may lead us in a superdeterministic direction} so we chose to trace out every degree of freedom with the exception of the pointer variable and the system. For the pointer variable we assume that the initial state is roughly the same for each run in the sense that there is a small positive number $\epsilon_{1}>0$ such that if $\vert r\rangle$ and $\vert r'\rangle$ are initial pointer states, then $\Vert\vert r\rangle-\vert r'\rangle\Vert\leq\epsilon_{1}$. This can be viewed as a weak version of an independence assumption between the measurement apparatus and the system. After tracing out the environment, we expect the time evolution operator to be different for different runs of the experiment. Assuming the flea model, the only difference is in the fleas. Since the hamiltonians for the different runs only differ in flea contributions, we may assume that there is another small positive number $\epsilon_{2}>0$, such that if $U$ and $U'$ are time-evolution operators describing, at the level of the system and the pointer, the change in state from the start of the measurement up to the end (including the introduction of the flea near the end), then $\Vert U-U'\Vert<\epsilon_{2}$. 

Using the previous assumptions we can derive an approximate version of (\ref{eq:fleabash}) without resorting to counterfactual reasoning. First, consider the following runs of the experiment:
\begin{equation*}
U_{i}\left(\vert m_{i}\rangle\otimes\vert r_{i}\rangle\right)=\vert m_{i}\rangle\otimes\vert E_{i}\rangle,
\end{equation*}
where the $\vert E_{i}\rangle$ are close to orthogonal, since these correspond to states with macroscopically distinct pointer values. Next, we perform a run where we prepared the superposition system state $\vert\phi\rangle=\frac{1}{\sqrt{2}}\left(\vert m_{1}\rangle+\vert m_{2}\rangle\right)$. From the estimates $\Vert U_{\phi}-U_{i}\Vert<\epsilon_{2}$, $\Vert\vert r_{\phi}\rangle-\vert r_{i}\rangle\Vert<\epsilon_{1}$ and linearity of the time-evolution operator we derive the inequality
\begin{equation} \label{eq:fleaorfight}
\Vert U_{\phi}\left(\vert\phi\rangle\otimes\vert r_{\phi}\rangle\right)-\frac{1}{\sqrt{2}}\left(\vert m_{1}\rangle\otimes\vert E_{1}\rangle+\vert m_{2}\rangle\otimes\vert E_{2}\rangle\right)\Vert<\sqrt{2}(\epsilon_{1}+\epsilon_{2}).
\end{equation}
Even though the post-measurement state is not equal to (\ref{eq:fleabash}), it is close enough to this state to conclude that tracing out the system will yield a non-pure state. By the previous reasoning, the wave function of the flea model cannot represent the state of the pointer variable in a straightforward way. But then, what does it actually describe? 

More pressingly; do asymptotic Bohrification and the flea model even deal with measurements? The flea model concentrates solely on collapsing a wave function, but much more is needed for any model of measurement. The post-measurement state not only needs to assign a value to the pointer variable, but also to the observable in such a way that these values are correlated. The flea model, concentrating only on the collapse of what we presume to be the pointer variable, is incomplete. Note that in Section~\ref{sec:mp} all incarnations of the problem of outcomes rely crucially on the existence of a correlation between observable and the pointer variable. So the flea model not only ignores an aspect which is integral to the very notion of measurement, but this aspect is also crucial to the problem that the flea model attempts to elucidate. 

It should be noted that \emph{we assumed} to apply the flea perturbation to solely try to collapse the \emph{pointer} to a definite outcome, which goes against the recommendations by Landsman himself. In his view, the collapse should occur on the level of the combined pointer and observable. Aside from the issues presented above, there is at least one other good reason for not wanting to identify the wave function with the pointer. To illustrate this, suppose we wish to include something akin to a flea perturbation in existing quantum measurement models. As an example, consider the model by Haake and Spehner \cite{hasp}, who use a double well for the pointer and an appropriate interaction to correlate the pointer's position with an observable (z-component of spin-1/2). The obvious place in the Haake-Spehner model to apply the flea is to the potential of the pointer. However, since the original hamiltonian of the model commutes with the observable, and the model is only modified at the level of the pointer, the Born probabilities associated to the observable are unaffected by the introduction of the flea. Even if the flea is effective in causing a collapse of the pointer wave function, the correlation between pointer and observable disappears. As a side note, introducing the flea model to the Haake-Spehner model is not as straightforward as one might think from the previous remarks as the model contains two separate potentials which are heavily-slanted double wells depending on the spin component, making it quite different from the flea model.

Thinking of entanglement and in keeping the observable/pointer correlation, we should consider Landsmans proposal to think of the flea model as representing the combined observable and pointer. However, to the authors it is not clear what is meant by this statement, and, more concretely, how to apply this idea to any physical model. In the case where the collapse occurs on the level of the pointer, it is intuitively clear how to construct a model, namely $x$ can, for example, describe the center-of-mass position of all particles in a gauge on the measurement device (pointer) that is subject to a electro-magnetic potential due to its interaction with the particles in the measurement device and the wavefunction is the center-of-mass wavefunction of the particles in the gauge. The flea can be imagined as a result of some change in the EM potential, although whether its cause should come from outside or inside the measurement device is unclear. But what if the collapsing wave function somehow represents the state of both pointer and observable? Where does the potential for this model come from, and how should it be interpreted. What would be the physical significance of $x$? What model can we build to give rise to a potential in $x$? It seems that the variable $x$ is now much harder to interpret. Without knowing what $x$ might stand for it is hard to answer questions such as: why is it energetically unfavourable to have large $x$ or $x=0$ for the pointer+observable? The values of $x$ are clear in the case of only a pointer (as in the Haake-Spehner model), but what does it mean in the case of a pointer and observable: does $x>0$ correspond to spin up and $x<0$ to spin down or only in the minima? What does $x=0$ mean for the observable? 

We end up in the situation where we do not have a physical interpretation for the potential, the wavefunction, or the flea, and where we are unable to connect with existing models of quantum measurement. Although this state of affairs is already troubling, it becomes more so after subsections \ref{sub:genfl}-\ref{sub:dyn2} when we conclude that the flea model is unable to perform its task of collapsing the wave function. How do you salvage a model if it is so incomplete at an interpretational level (i.e., what does the wave function mean, where does the flea come from, what does the potential represent) and at the same time so far removed from any other model of quantum measurement (it would need a model containing a bound state representing the combined observable and pointer state and which is completely quantum mechanical).

\subsection{Generalisations} \label{sub:genfl}

We consider two generalisations of the flea proposal; measurements with more than 2 possible outcomes, and system preparations with unequal associated Born probabilities. We could have chosen to generalise in a different direction. For example, we could trade in the Hilbert space $L^{2}(\mathbb{R})$, the hamiltonian with a double well potential, and the semi-classical limit $\hbar\to0$ in for the spin chain Hilbert space $\mathcal{H}_{N}=\otimes^{N}\mathbb{C}^{2}$, the quantum Curie-Weisz hamiltonian, 
\begin{equation*}
H_{N}=-\sum_{-\frac{N}{2}}^{\frac{N}{2}-1}\sigma^{z}_{i}\sigma^{z}_{i+1}-B\sum_{i=1}^{N}\sigma^{z}_{i}
\end{equation*}
and the limit $N\to\infty$, where $N$ is the number of sites on the lattice. This example was treated in the setting of asymptotic  Bohrification in \cite{landsman}. The only thing that we need to add is a flea, for example in the shape of a matrix $W=\text{diag}(0,0,\ldots,0,\epsilon,0,\ldots,0)$, where $\epsilon$ is some small number and the matrix representation is relative to the basis of $\mathcal{H}_{N}$ generated by the eigenstates of the operators $\sigma^{z}_{i}$. However, because of length issues, we do not discuss such examples any further, at the risk of giving the impression that the idea of the flea model is limited to $n$-well potentials and the semi-classical limit.

\subsubsection{More outcomes} 

\begin{figure}
\begin{centering}
\includegraphics[scale=0.35]{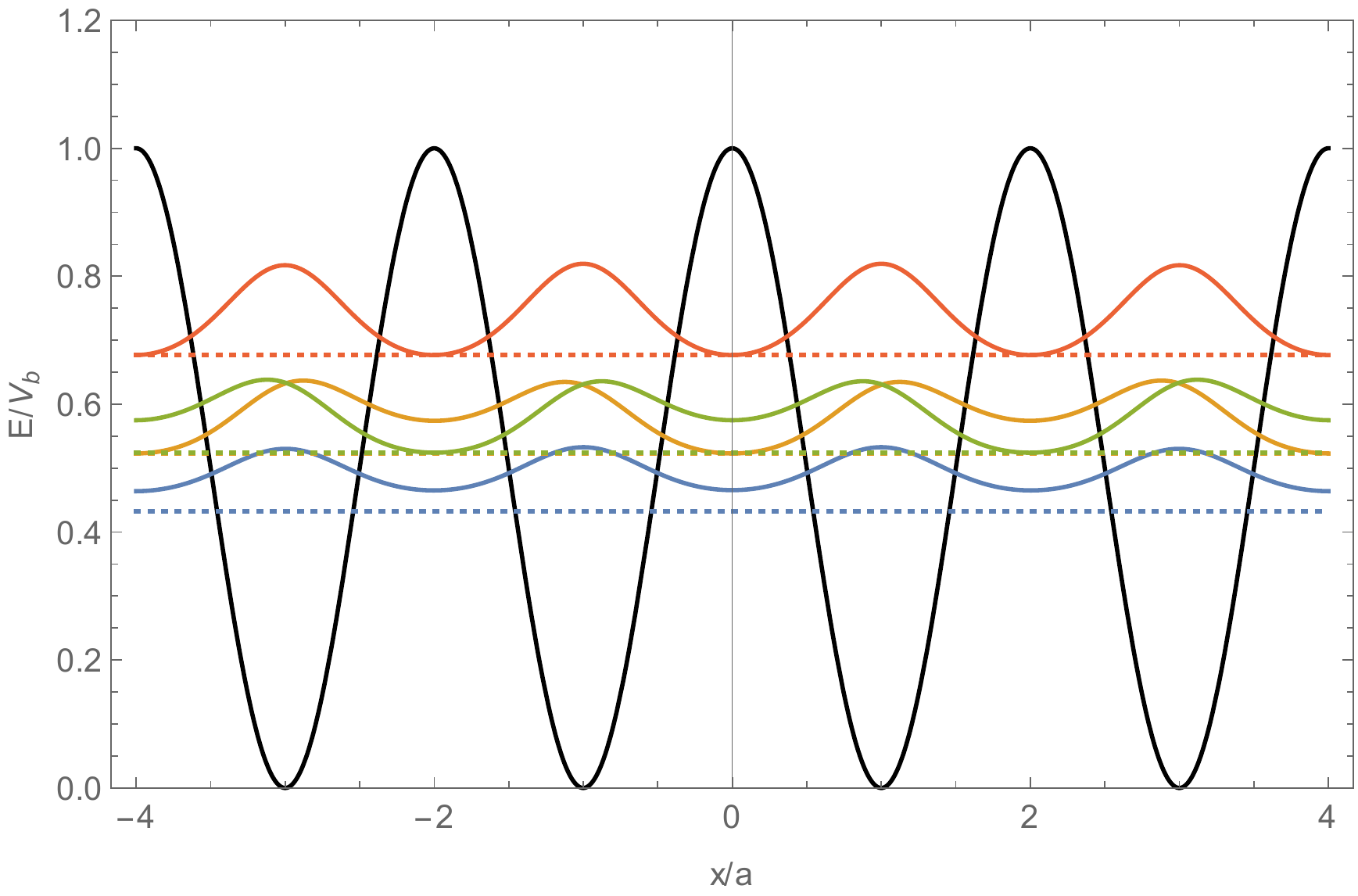}\includegraphics[scale=0.35]{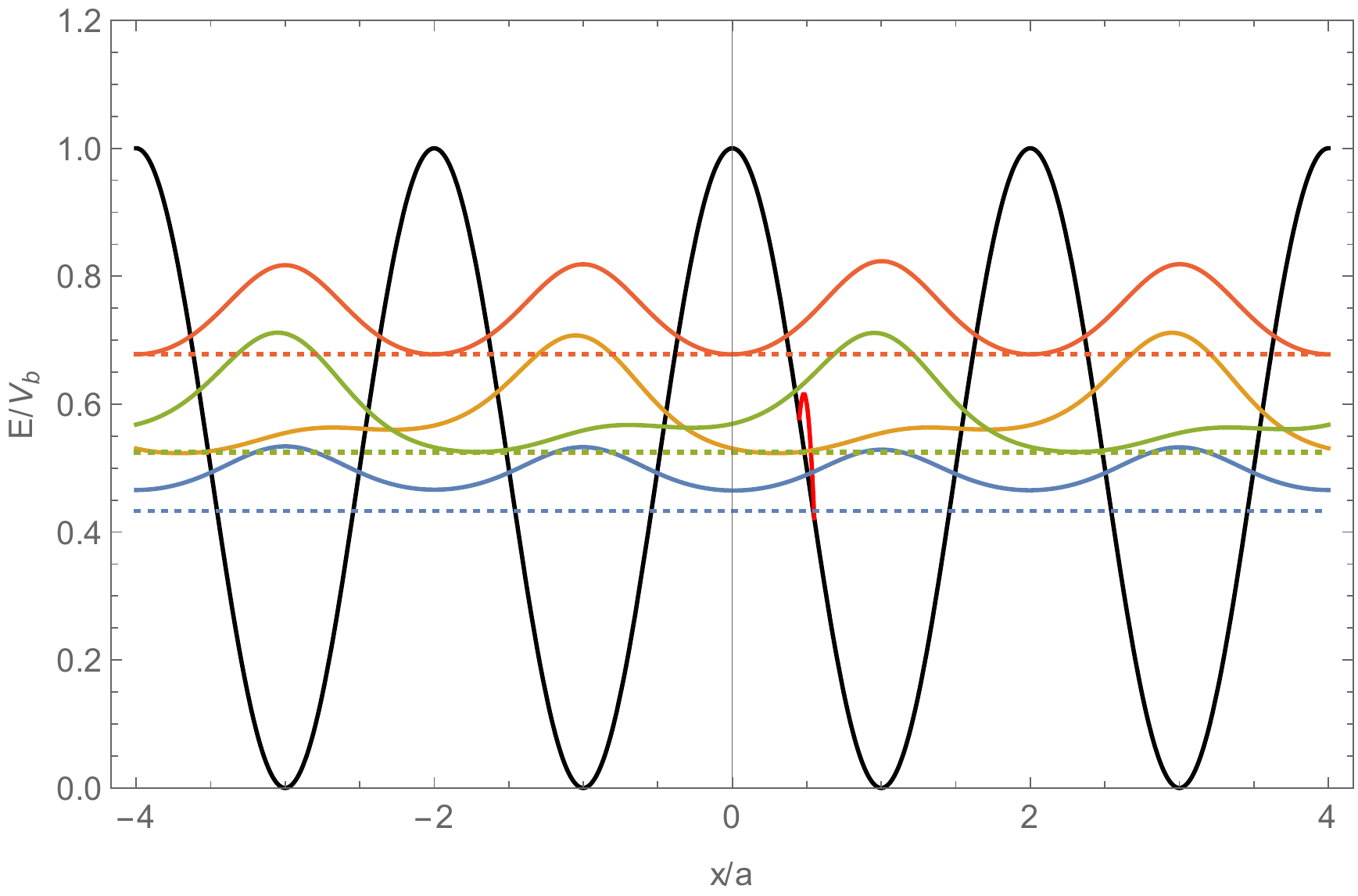} 
\par\end{centering}

\begin{centering}
\includegraphics[scale=0.35]{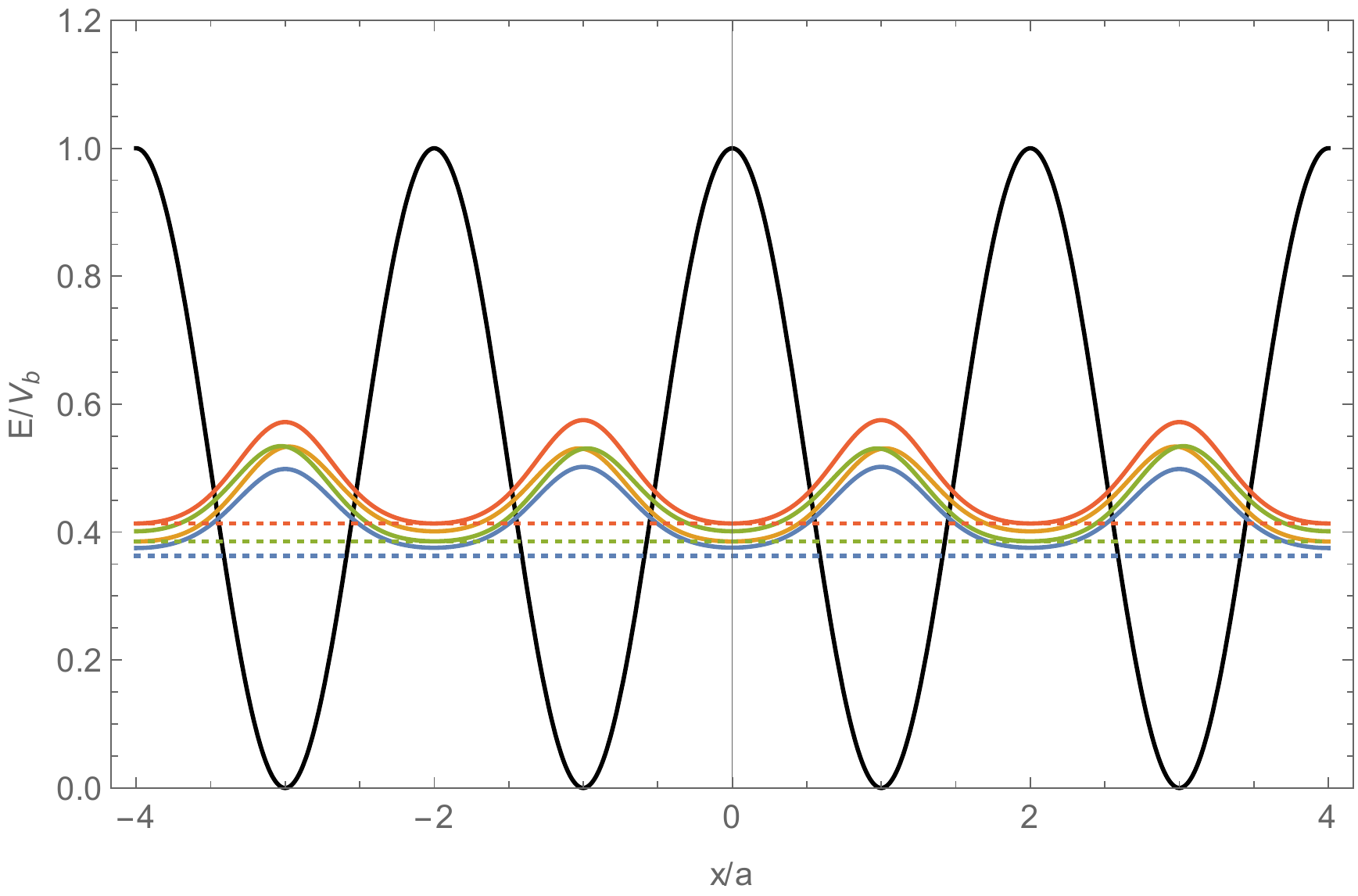}\includegraphics[scale=0.35]{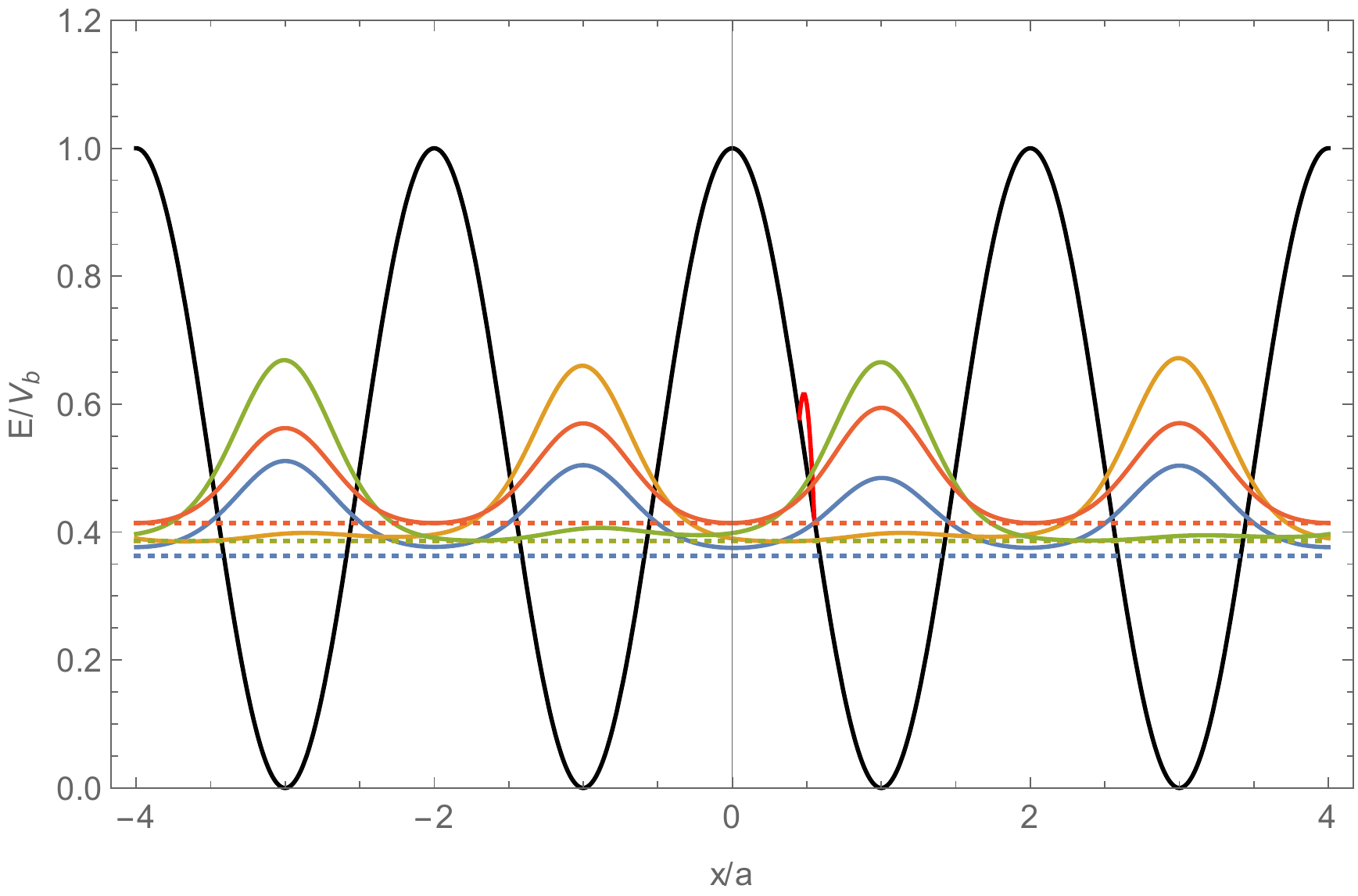} 
\par\end{centering}

\begin{centering}
\includegraphics[scale=0.35]{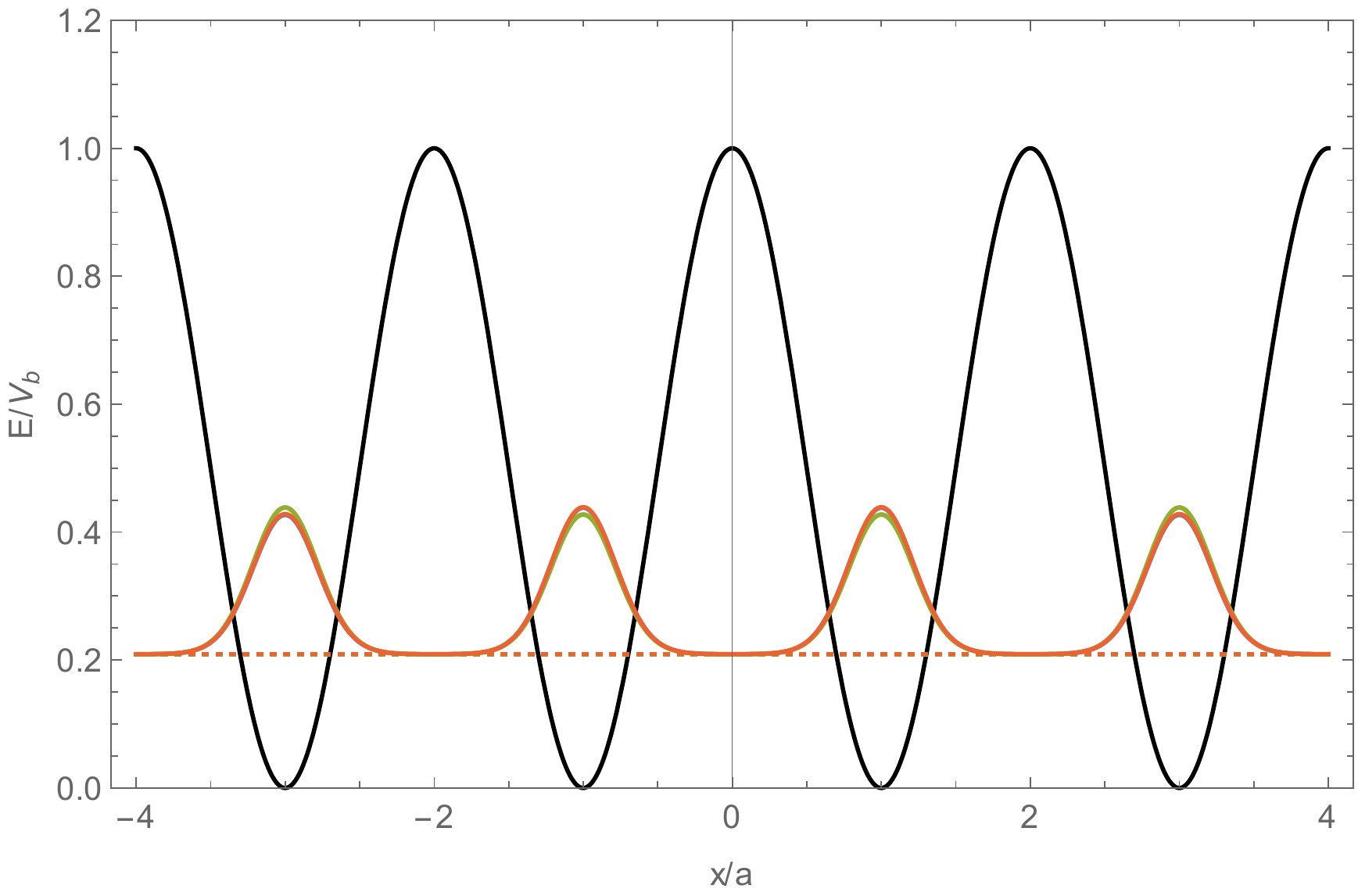}\includegraphics[scale=0.35]{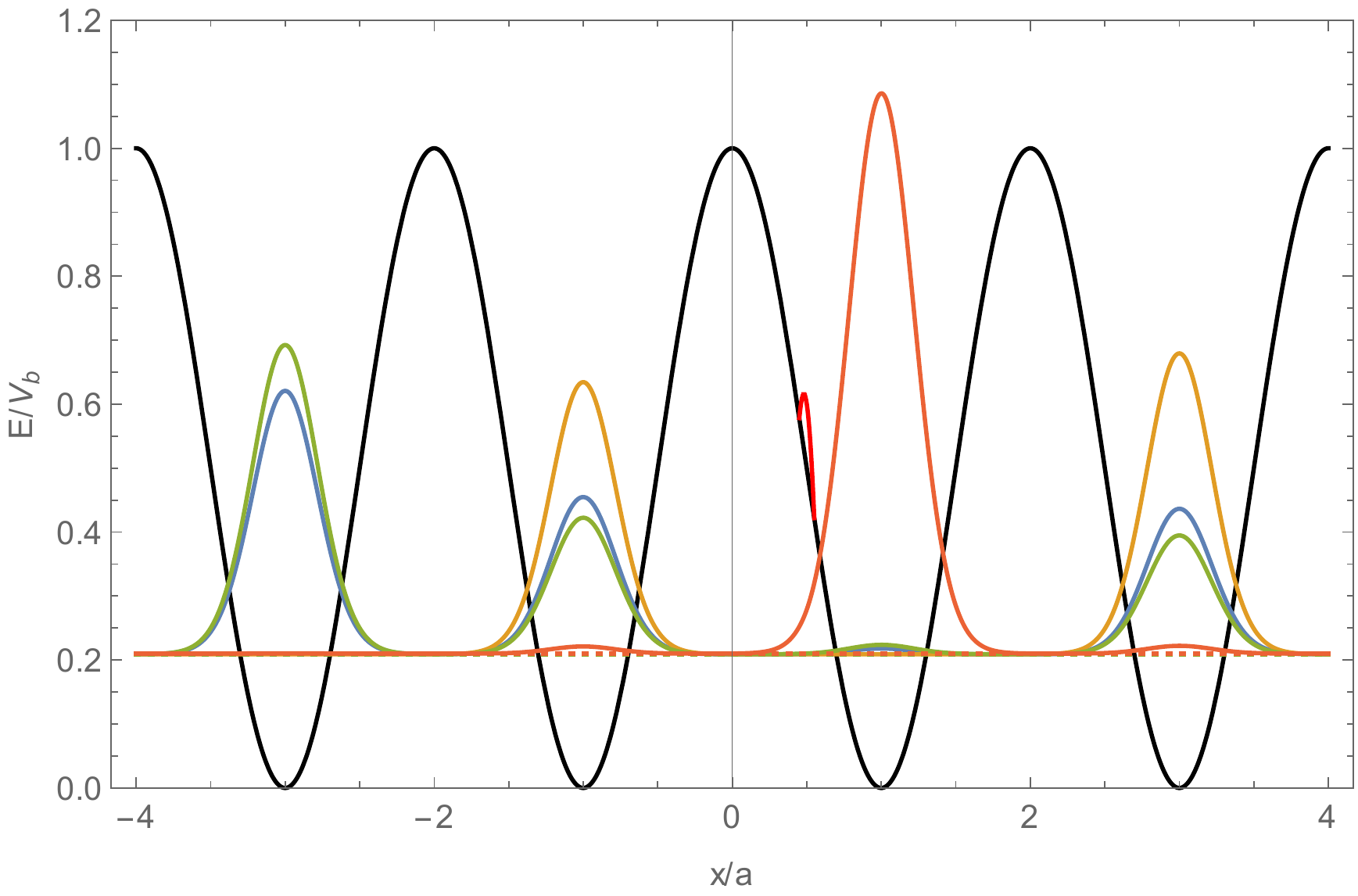} 
\par\end{centering}

\protect\protect\caption{The first four eigenfunctions as $\left|\psi(x)\right|^{2}$ of the
periodic potential without (left) and with (right) a flea for $\xi=\left\{ 0.6,0.4,0.2\right\} $
from top to bottom, respectively. Clearly, without a flea the solutions are symmetric and will remain present in all wells as $\xi$ becomes smaller. With the flea the symmetry is broken, and the groundstate (blue) will move out from the well with the flea.} \label{fig:Periodic_Pot}
\end{figure}

The two wells of the potential correspond to the two values of the observable. How should we generalise the model to observables with $n$ distinct values, where $n>2$? One strategy is to consider the groundstate of an $n$-well potential with periodic boundaries.
As an example we use $V(x)=V_{b}\cos^{2}\left(\pi x/2a\right)$ on $x\in\left[-na,na\right]$. This choice has the advantage that energy eigenstates are known, namely they are the Mathieu functions, shown in Figure~\ref{fig:Periodic_Pot}. 

A flea perturbation is added to the potential and the eigenfunctions are determined numerically. From Figure~\ref{fig:Periodic_Pot} it
is seen that the groundstate need not localize in a single well when only a single flea perturbation is added. In the figure, the
flea perturbation is parabolic with finite support, although these details have no impact on the discussion. Since a single flea does not provide a fully localised ground state, we should consider adding multiple flea perturbations in the different wells. These multiple fleas must be chosen in such a way that the resulting potential no longer has any symmetry. If many fleas are added, and their locations, sizes and shapes are chosen largely at random, then this should be no problem. 

A potential cause of problems for this generalisation is the increase in collapse times, especially when there are many wells, providing many barriers through which parts of the wave function have to tunnel. In Subsection~\ref{sub:dyn2} it will become clear that collapse times are problematic for the double well setting already. Therefore we shall not explore the increase of collapse times due to the added wells.

\begin{figure}
\begin{centering}
\includegraphics[scale=0.3]{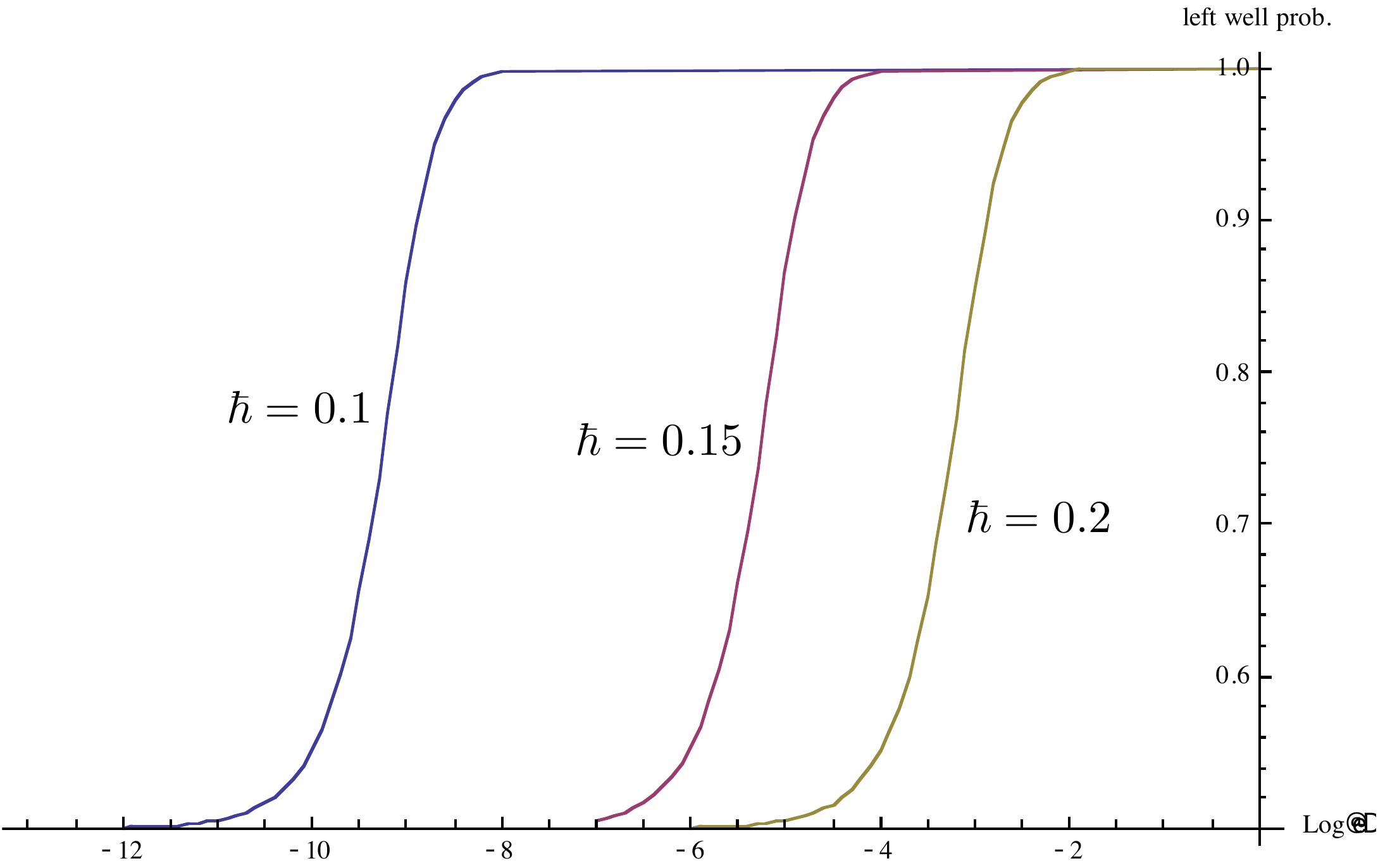} 
\protect\caption{Left-well probability of the groundstate of a double-well potential as a function of flea size (shrinking parameter $\epsilon$), for $m=1$ and $\hbar=\xi\in\{0.1,0.15,0.2\}$. The shrinking parameter $\epsilon$ is scaled logarithmically with base 10. This shows that for a fixed flea size there is some $\xi$ below which the flea causes the groundstate to be almost completely located in a single well. However, it also shows that adding an asymmetry (interpreted as a flea) to the double well to allow for unequal initial probabilities relies critically on the size of the asymmetry.} \label{fig:uneq_born}
\end{centering}
\end{figure}

\subsubsection{Unequal Born Probabilities}

How can we adapt the flea model to deal with other Born probabilities than the 50/50 case? At least two options are available. The first option is to keep using a symmetric potential, but to add additional wells. Suppose we consider a two level system, and that the Born probability assigned to one of the two values of the observable can be expressed as a rational number $p/q$, where we assume that 1 is the only common divisor of natural numbers $p$ and $q$. We can consider $q$ wells, $p$ of which correspond to the value with the Born probability $p/q$, and the other $q-p$ correspond to the other value. Then we can proceed as before. This option will typically involve many wells, making it complicated and providing potential additional problems with regards to collapse times. In addition, this option looks far removed from the idea that the system state becomes correlated to the pointer variable. Since the connection between observable and pointer is already so weak in the flea model, we consider this to be a serious drawback.

A second option, which does not share the previous drawbacks, is to replace the symmetric ground state of the symmetric double well potential by an asymmetric potential yielding a ground state such that the probabilities assigned to the wells are equal to the desired Born probabilities. However, since we are working in the semi-classical limit, the asymmetry of the potential should shrink with the value of $\hbar$. Otherwise, the asymmetry localises the wave function, just as the flea would.

In addition to the symmetric potential $V$ there are three additional asymmetric terms added to the potential:
\begin{equation}
V(x)+W_{0}(x)+W_{b}(x)+W_{f}(t).
\end{equation}
There is some noise $W_{0}$ which is too small (in sup-norm) to affect localisation, and which may be time dependent. We only add it to emphasise that the initial potential need not be completely symmetric in order to start out with a (sufficiently) symmetric ground state. Otherwise, we would need to worry whether the flea model itself conflicts with Earman's principle. Then there is a contribution $W_{b}$ which ensures that the initial state of the flea model has the desired Born probabilities. Since it affects the Born probabilities, it is larger (in sup-norm) than the noise, but must be smaller than the flea in order to avoid localisation. The final contribution is the flea $W_{f}(t)$ which localises the wave function. As an example consider Figure~\ref{fig:uneq_born} where, for three different values of $\hbar$, we consider a parabolic flea $W$. We shrink this flea as $\epsilon W$, where $\epsilon$ is a number ranging from $10^{-1}$ to $10^{-12}$. As the flea shrinks we consider the probability assigned to the left well. The sensitivity of the ground state relative to small perturbations is seen as the curves shift to the right, when $\hbar$ decreases. 

Suppose we are given a value of $\hbar$ and some perturbation $W$. If $W$ is to be part of the noise, then there is an upper bound such that if we shrink $W$ below this bound, then it will not affect the Born probabilities. Likewise, if $W$ is intended as the flea, then there is a lower bound, and if $W$ is, relative to the supnorm, larger than this boundary, it will allow localisation. If $W$ is needed to set the initial Born probabilities, then there is only a single value of $\epsilon_{0}$, such that $\epsilon_{0} W$ provides the right Born probabilities. Any deviation from this value affects the Born probabilities. Note that the variations around $\epsilon_{0}$, such that the Born probabilities do not change significantly, decrease in order of magnitude as $\hbar$ decreases. More importantly, in all cases these asymmetries that we are fine-tuning are necessarily smaller than the flea perturbation. The flea is supposed to represent the result of an extremely small environmental fluctuation. Yet, we are fine-tuning even smaller fluctuations, just to get the right initial state. 

In addition to the previous fine-tuning problem, the problem of independence also plays a role here. It is the location of the flea that determines in which well the final state is localised! The Born probabilities, modeled using a finely tuned asymmetry, do not matter when a flea is chosen. In \cite{lare}, the flea perturbation was compared to a hung parliament, where a small political party acquired influence far exceeding its relative size. We feel that the flea is more like a dictator, determining what is going to happen, unfettered by any democratic constraint. The Born probabilities of the initial wave function, motivated by the correlation between observable and pointer variable, do not play any active role in the model.

We can only conclude that none of the options presented here provides a satisfactory extension of the flea model to arbitrary Born probabilities. But of how much interest can a solution to the problem of outcomes be, if it can only be applied to the special 50/50 case?

\subsection{Problem of Dynamics} \label{sub:dyn1}

We ask to what extent the problem of outcomes, as formulated in Subsection~\ref{sub:poo1} applies to the flea model. The flea model concentrates on a single degree of freedom which is related to the pointer variable $\mathcal{P}$. As discussed in the previous subsection, the precise relation is unclear, but it is clear that the different potential wells correspond to different values of the pointer variable. We assume that there is a two-level system $\mathcal{S}$ and that initially the system state is the superposition
\begin{equation}
\vert\psi\rangle=\frac{1}{\sqrt{2}}\left(\vert m_{1}\rangle+\vert m_{2}\rangle\right),
\end{equation}
of distinct eigenstates of the non-degenerate observable. It is assumed that the eigenstates become correlated to the pointer variable. As a result, the superposition of the system state is transferred to a superposition for the pointer variable, a Schr\"odinger cat state. Next, we fix an environment $\mathcal{E}$, including the pointer variable, and any degree of freedom relevant to the measurement. This includes all degrees of freedom that are involved in the creation of the flea. For such a large enough environment we can assume that the time evolution proceeds through a unitary operator $U$. Otherwise, contrary to the agenda of the flea model, we would not be considering a collapse within the formalism of quantum theory.\\

If the flea is effective, then the post-measurement state, reduced to the pointer variable, is concentrated in only one of the two potential wells. At this point, the pointer variable needs to be correlated to the observable. Hence, the post-measurement state, reduced to the system, is concentrated around one of the two eigenvalues of the observable. Clearly, the Born probabilities associated to the eigenvalues of the observable change significantly during the measurement.

As in Subsection~\ref{sub:poo1} we can express the unitary operator as a matrix of operators acting on $\mathcal{H}_{\mathcal{E}}$.
\begin{equation*}
U = \left(
\begin{array}{cc}
U_{11} & U_{12} \\
U_{21} & U_{22}
\end{array} \right),
\end{equation*}
From the same subsection we know that if there exists a positive number $\epsilon\ll1$, such that the operators $U_{12}$ and $U_{21}$ satisfy $max\{\Vert U_{12}\Vert,\Vert U_{21}\Vert\}\leq\epsilon$ with respect to the operator norm, then the Born probabilities associated to the observable vary very little. In other words, if the operator $U$ is almost diagonal, then the flea cannot cause a collapse in the sense described above.\\

Let $\rho_{i}$ be a suitable initial state, such that $Tr_{\mathcal{E}}(\rho_{i})$ is close (in norm) to $\vert m_{i}\rangle\langle m_{i}\vert$, then for the evolved state $U^{\dagger}\rho_{i} U$ the system state $Tr_{\mathcal{E}}(U^{\dagger}\rho_{i} U)$ remains close to $\vert m_{i}\rangle\langle m_{i}\vert$. But since we do not allow for counterfactual reasoning we cannot, as argued in Subsection~\ref{sub:poo2}, invoke linearity to conclude that $U$ is approximately diagonal. The reason is that we only consider states, that are suitable as initial states of a measurement run. Even if $\rho_{1}$ and $\rho_{2}$ are two suitable initial states, non-trivial convex combinations thereof need not correspond to a suitable initial state for an experiment. It is the possibility of superselection rules for initial states of the measurement that prohibits us from deriving an approximate diagonal form of $U$. Without such rules, either the flea is not effective, or it is not of quantum mechanical origin. 

The need for superselection rules is undesirable. If the environment $\mathcal{E}$ is very large, including the preparation setup for $\mathcal{S}$, then indeed we expect different Born probabilities associated to the observable to correspond to different environmental states, since these correspond to different setups of the preparation apparatus. But dragging the preparation devices into the discussion brings us closer to the position of superdeterminism. The very notion of measurement carries a sense of independence between system and measuring device with it. This is typically expressed through an initial state which is factorised between system and the rest. Defending superselection rules for initial states may violate this notion of independence up to the point that the reader may think that we are no longer concerned with a realistic model for a measurement. This same issue holds for the problem of independence from Subsection~\ref{sub:dis3} as well.

\subsection{Problem of Collapse Times}  \label{sub:dyn2}

Under the assumptions of the previous subsection, a dynamical collapse mechanism of the wave function within quantum mechanics was ruled out. However, it is not completely clear whether the assumption of an almost diagonal unitary operator $U$ applies to the flea model. The fact that the flea model is silent about the correlations between the system observable and the pointer variable does not help in answering this question either. Consider the following two arguments which cast doubt on the applicability of the previous result:
\begin{enumerate}
\item The flea model only considers system state preparations yielding states with fifty-fifty Born probabilities for the observable. As argued in Subsection~\ref{sub:genfl}, generalising the model to states with other Born probabilities is non-trivial at best. Preparations with different associated Born probabilities are needed in deriving the almost diagonal form of $U$.  
\item If we seek to solve the problem of outcomes within quantum mechanics, then we should take the problem of independence into account. This means that we should not a priori ignore entanglement of the system with environmental degrees of freedom. Dependencies between the reduced system state and the reduced environmental state also prohibit us from deriving an almost diagonal form for $U$ by invoking linearity of the time evolution. 
\end{enumerate}

These arguments may strike the reader as weak, since the first argument is based on the inability of the flea model in considering arbitrary system preparations, and with regards to the second argument the flea model itself does not deal with the problem of dependence as it does not postulate a relation between the flea and the Born probabilities of the system state. Even so, if we do not discount the possibility that these problem can be overcome, then such a possible solution may render the no-go result of the previous subsection irrelevant to the flea proposal. Therefore, in this subsection, we consider the time scales of the collapse for the flea model without making assumptions on $U$. For a macroscopic device, modelled as a small value for $\hbar$, we find that a collapse takes an unrealistically long time. Although this result agrees with our findings of the previous section, the reasoning is different. In Subsection~\ref{sub:dyn1}, the Born probabilities of the observable, and correspondingly also the Born probabilities of the correlated pointer variable, barely change over time because of the existence of an approximate measurement basis. In the current subsection the Born probabilities of the pointer variable barely change over time because we are considering a tunnelling problem with respect to a relatively large potential barrier.\\ 

Recall the work on Schr\"odinger operators by Jona-Lasinio et.al.~\cite{jola}, which plays a key role in the flea approach. In the words of Landsman and Reuvers \cite{lare}: 
\begin{quote}
...the ground state of a symmetric double-well Hamiltonian (which
is paradigmatically of Schrodinger's Cat type) becomes exponentially
sensitive to tiny perturbations of the potential as $\hbar\to0$. 
\end{quote}
This phenomenon may have relevance for the quantum to classical transition. However, being a relation between bound states for slightly different potentials, it is a \emph{static} phenomenon that has no obvious connection to the \emph{dynamical} problem of the collapse of the wave function. In fact, as we argue below for the double well, when we decrease $\hbar$, we are effectively increasing the potential barrier between the wells, and the time needed for symmetric wave function to localise increases at least exponentially in the semi-classical limit, regardless of the way in which the flea is introduced. Since smaller values of $\hbar$ are supposed to represent a larger degree of macroscopicity, such an increase in collapse times invalidates the flea model.\\

How should we choose the time dependence of the flea? As there is presently no physics explaining the emergence of the flea we treat this to a large extent as a mathematical problem, rather than a physical one. The only physical restriction is that the collapse times should decrease as the scale $\hbar$ decreases, since this parameter is used to represent the degree of macroscopicity. Landsman and Reuvers \cite{lare} consider various ways of dynamically introducing a flea term $W(x)$ to the potential. The first option is to introduce it as a `quench', 
\begin{equation}
H(t)=H_{0}+\theta(t)\epsilon W,
\end{equation}
where $\theta(t)$ is the Heaviside step function, and $\epsilon>0$ is a real number. Depending on the size of $\epsilon$, introducing the flea either results in a wave function which oscillates between the two wells, or the original symmetric wave function barely changes at all. For no value of $\epsilon$ however, do we obtain a wave function which is localised in a single well. 

\begin{wrapfigure}{r}{0.45\textwidth}
\includegraphics[scale=0.6]{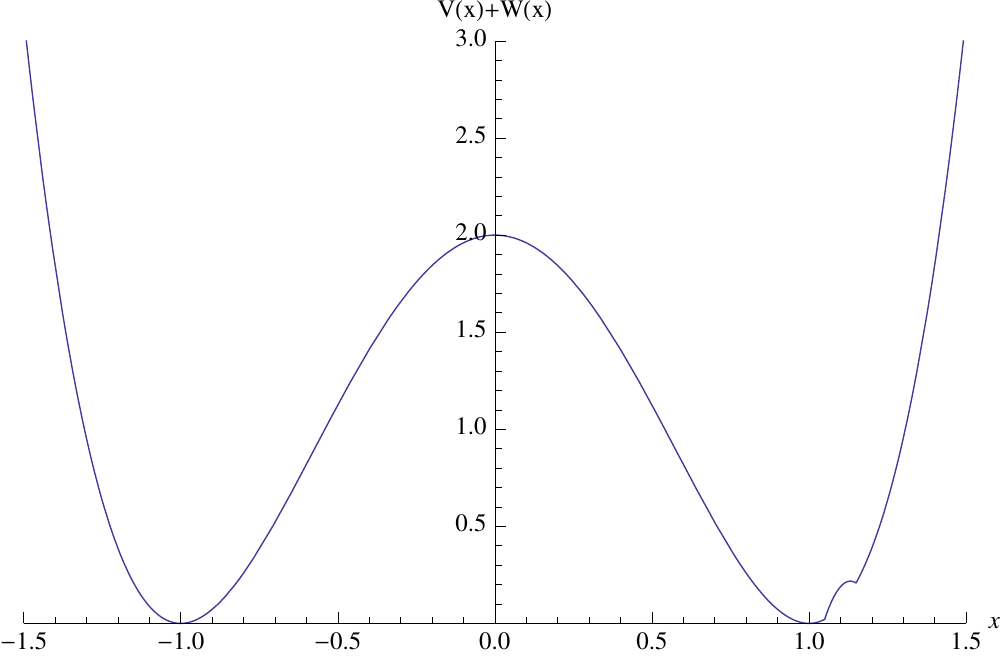} 
\protect\caption{Potential with Flea \label{fig:potential}}
\end{wrapfigure}

Other attempts, such as adding white noise or Poisson noise did not lead to a localisation either. In all these attempts both the ground state and the first excited state of the perturbed setting contribute significantly to the wave function. In order to avoid this, Landsman and Reuvers consider introducing the flea in the adiabatic limit, ensuring that we end up with the ground state of the perturbed setting.
As it turns out, the combination of the adiabatic limit in quantum mechanics and the semi-classical limit poses a problem of too large collapse times. For the sake of concreteness, consider the time-dependent Hamiltonian: 
\begin{equation}
H(t)=H(0)+\sin\left(\frac{\pi t}{2T}\right)W,\ \ \text{for}\ \ t\leq T
\end{equation}
and $H(t)=H(0)+W$ for $t>T$. For the potential $V$ of $H(0)$ we use the symmetric double well potential as in (\ref{equ_hamop}) and the flea $W$ is a parabolic shape, as shown in Figure~\ref{fig:potential}.

Let $(\psi_{n})_{n\in\mathbb{N}}$ be a (time-dependent) orthonormal basis of eigenfunctions of $H(t)$, where the eigenvalues $E_{n}(t)$ are ordered in increasing size. The wave function can be expressed as: 
\begin{equation}
\Psi(t)=\sum_{i=1}^{\infty}c_{n}(t)\psi_{n}(t)\exp\left(\frac{i}{\hbar}\int_{0}^{t}dsE_{n}(s)\right),
\end{equation}
where the $x$-dependence is suppressed in the notation. We start out in the ground state of $H(0)$, so $c_{0}(0)=1$ and
$c_{n}(0)=0$ for all $n>0$. The time dependence of $c_{n}(t)$ is
given by 
\begin{equation}
\dot{c}_{n}=c_{n}\langle\psi_{n}\vert\dot{\psi}_{n}\rangle-\sum_{m\neq n}c_{m}\frac{\langle\psi_{n}\vert\dot{H}\vert\psi_{m}\rangle}{E_{m}-E_{n}}\exp\left(-\frac{i}{\hbar}\theta_{mn}(t)\right),
\end{equation}
\begin{equation}
\theta_{mn}(t)=\int_{0}^{t}ds(E_{m}(s)-E_{n}(s)).
\end{equation}
In what follows we concentrate on $\vert\dot{c}_{1}(0)\vert$ for several reasons. As already noted, the first excited state is localised in the other well with respect to the ground state, therefore $\vert\dot{c}_{1}(t)\vert$ is of interest. In addition, consider the energy splitting $\Delta(t)=\vert E_{0}(t)-E_{1}(t)\vert$, which rapidly decreases in the semi-classical limit. During the collapse, the splitting $\Delta(t)$ takes on its smallest value for the unperturbed setting $t=0$. It is also at this time that we expect the overlap $\vert\langle\psi_{1}(t)\vert W\vert\psi_{0}(t)\rangle\vert$, appearing in the matrix element $\vert\langle\psi_{1}(t)\vert \dot{H}\vert\psi_{0}(t)\rangle\vert$, to be at its largest. In other words, if $\vert\dot{c}_{1}(0)\vert$ turns out to be sufficiently small, then this may help yield a final state which is close to the ground state of the perturbed potential. Thus, we ask for which order of magnitude of $T$, does the quantity 
\begin{wrapfigure}{r}{0.4\textwidth}
\includegraphics[scale=0.7]{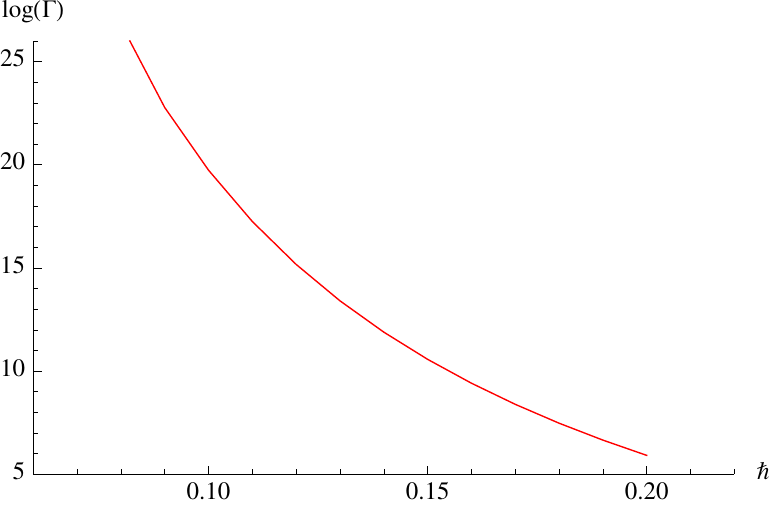} 
\protect\caption{Collapse times \label{fig:collapsetime}}
\end{wrapfigure}
\begin{equation}
\vert\dot{c}_{1}(0)\vert=\frac{\pi}{2T\Delta(0)}\vert\langle\psi_{1}(0)\vert W\vert\psi_{0}(0)\rangle\vert\label{fleadyntrubble}
\end{equation}
become sufficiently small. As in~\cite{lare}, as $\hbar\to0$,
$\Delta(0)$ decreases as 
\begin{equation}
\Delta(0)\approx\frac{2\hbar a\sqrt{\lambda}}{\sqrt{e\pi}}e^{-\frac{d_{V}}{\hbar}}.
\end{equation}

Unless $\vert\langle\psi_{1}(0)\vert W\vert\psi_{0}(0)\rangle\vert$ decreases with a rate of at least $e^{-d_{V}/\hbar}$ in the semi-classical limit, the time $T\to+\infty$ must increase exponentially fast in that same limit, in order to keep $T\Delta(0)$ finite. Here $d_{V}$ denotes the WKB-factor. In itself, the limit $T\to+\infty$ as $\hbar\to0$ need not be problematic. As $\hbar$ decreases, we can shrink the flea $W$, and consequently shrink $\vert\langle\psi_{1}(0)\vert W\vert\psi_{0}(0)\rangle\vert$, whilst retaining a localised perturbed ground state. Consider the quantity:
\begin{equation}
\Gamma=\frac{\vert\langle\psi_{1}(0)\vert W\vert\psi_{0}(0)\rangle\vert}{\Delta(0)},
\end{equation}
which we use to quantify the rate at which $T$ needs to increase. Figure~\ref{fig:collapsetime} shows how $\text{Log}(\Gamma)$ varies with $\hbar$ for the fixed parabolic flea of Figure~\ref{fig:potential}. Note the exponential growth, indicating that $T$ needs to increase at a rate $x\mapsto e^{e^{x}}$. In the range of the graph, if we halve the value of $\hbar$, then $\Gamma$ increases roughly 15 orders of magnitude. How much can we shrink the flea $W(x)\mapsto10^{-n}W(x)$ in order to compensate for this effect. For any $\hbar$ in the range of Figure~\ref{fig:collapsetime}, if $n\geq12$, the perturbed ground state is no longer localised in a single well. In other words, the collapse times rapidly increase as $\hbar\to0$ even if we shrink the flea $W$ as much as possible along the way. Why do we expect this behaviour to hold in general? Instead of $\hbar$, consider $\lambda:=1/\hbar$. Regardless of the details of the flea, we are considering situations where part of the wave function has to tunnel through a barrier of increasing height $\lambda V_{0}$. 

If the flea is introduced too fast, then the first excited state becomes occupied and we retain the superposition. If the flea is introduced slower, to ensure that we remain in the ground state, then the collapse times increase dramatically as the macroscopicity scale parameter $\hbar$ decreases. The cause of the increase of collapse times is the rapid decrease of the initial splitting $\Delta(0)$. Yet, this same smallness of the splitting with regards to the semi-classical limit is important to the asymptotic Bohrification programme, as argued in~\cite{landsman}:
\begin{quote}
The essential point is that in our models, the energy difference $\Delta E_{\bullet}=E_{\bullet}^{(1)}-E_{\bullet}^{(0)}$,
between $\Psi_{\bullet}^{(1)}$ and $\Psi_{\bullet}^{(0)}$ vanishes
exponentially as $\Delta E_{N}\sim exp(-C\cdot N)$ for $N\to\infty$,
or as $\Delta E_{\hbar}\sim exp(-C'/\hbar)$ for $\hbar\to0$ respectively.
This means that asymptotically any linear combination of $\Psi_{\bullet}^{(1)}$
and $\Psi_{\bullet}^{(0)}$ is almost an energy eigenstate. 
\end{quote}

The flea model hinges on the precept that first a macroscopic superposition for the pointer variable is formed, and subsequently this superposition is broken down dynamically. As argued in this subsection, this is an open invitation to problems with regards to tunnelling times. However, conceptually it is suspicious as well. Already from decoherence we know that setting up a macroscopic superposition is highly non-trivial. In the next, and final, subsection we explain that a true dynamical treatment of the flea model conflicts with its own foundation.

\subsection{The problem with potentials}

The flea proposal is based on the observation that two different potentials, one with a flea and the other without, behave quite differently when the parameter $\xi$ is changed. This is, however, a completely static argument. The proposal relies heavily on potentials, which according to quantum theory are themselves always to be seen as an approximation of some dynamic quantum fields. For example, the possible energy of a photon emitted by an atom can be understood by the energy levels of the electro-magnetic potential governing the electrons, however, to understand the dynamics of the emission the electrons must be coupled with an electro-magnetic field. This is because the dynamics depends not only on the system's state but also on the state of the environment, in this case the EM-field, whose interplay determines the speed of the changes in the system. This means that to connect the two different potentials, the flea must be dynamically added and thus it must be treated as a dynamic quantum field. 

Without an interpretation of the flea, we tried in the previous subsection the only possible way to gain some insight into the dynamics by applying a flea adiabatically, which by the very nature of adiabatic processes gives, not surprisingly, slow collapse times. The reason why the process needed to be adiabatic, is that otherwise the model would not stay in the groundstate and thus not lead to the required final state. In conclusion, this subsection serves as a reminder that \emph{\bf{the measurement problem is a dynamical problem}}. Since the idea behind the flea model relies heavily on potentials, it is hard to envisage what would remain of it in a full dynamical quantum treatment.

\section{Discussion}

In its current form, the flea model is not feasible as a model for dynamical collapse of the wave function within the formalism of quantum theory.
We briefly list the reasons for our conclusion. Some of these are shared with other attempts at solving the measurement problem, such as:
\begin{itemize}
\item The problem of defining outcomes: The notion of outcome relies on pure states to define certainty within quantum theory (ground state).
\item The problem of epsilonics: There remains a small contribution to the other outcome, except in the idealised limit.
\item The problem of independence: It is unclear how the flea connects to the system state to reproduce the Born rule. Location of the flea solely determines outcome.
\end{itemize}
Also some of our objections are particular to the flea proposal:
\begin{itemize}
\item Problem of potentials: The idea behind the flea proposal comes from comparing two separate potentials and thus has no bearing on dynamics. True dynamics only follows from a treatment in terms of two coupled quantum systems.
\item Physical interpretational issues: The origin of both the potential and flea is unclear, which makes a realistic physical model almost impossible.
\item Adiabatic introduction of the flea gives slow collapse times with adiabatic flea. Other modes of introducing the flea, investigated thus far, do not provide a collapse at all.
\item The inability to generalise to system states with unequal associated Born probabilities.
\item It is unclear how to scale to more than two outcomes.
\item The pointer-observable correlation becomes broken if the flea model is interpretated at the level of the pointer.
\end{itemize}

One major problem is that both the flea model and asymptotic Bohrification are silent on how the interaction between the observable and the pointer gives rise to the double well symmetric ground state. Correlating the pointer and the observable lies at the very heart of the notion of quantum measurement, and any treatment of the problem of outcomes that is not explicit on this aspect is incomplete. Even if the model could be applied to the pointer+observable to re-establish the correlation, then the model still rests on the Born rule to provide an interpretation of the outcome at finite $\hbar$ (the problem of epsilonics). Furthermore, the Born rule must be used to provide a physical interpretation to the model by connecting the variations of fleas to the statistics set by the initial state (the problem of independence). 

If all these points are ignored in an attempt to solve only the problem of outcomes, the fact remains that the flea model by itself provides no physical interpretation of the potential or of the flea. Without a physical interpretation we can only use the potential model, which is insufficient to determine the dynamics. In lieu of a real dynamical model, we demonstrated in Subsection~\ref{sub:dyn1} that the existence of an approximate measurement basis already leads to unrealistically long collapse times if an adiabatic process is assumed, which is done to artificially keep the system in the groundstate at all times, unless additional selection rules for the initial states are postulated.
 In addition, the environment is only mentioned as the cause of the flea, but it does not play an explicit role. So although the flea model is fully quantum mechanical in the sense that it works within the operator algebraic setting of formal quantum mechanics, it is not so clear that it is fully quantum mechanical in the sense that there is a quantum-mechanical model of measurement, which reduces to the flea model after tracing out all the environmental degrees of freedom. For the reader who is not about to give up on finding a solution to the problem of outcomes along the lines of asymptotic Bohrification; finding such a model with suitable environment should be the number one priority. Providing physical grounding of the flea may help with the problem of independence and provide some guidance on how to proceed further, as straightforwardly applying the sensitivity of bound states to a flea perturbation does not work.

Alternatively, the reader may conclude that the encountered problems indicate that the problem of outcomes cannot be overcome within unitary quantum mechanics. The problem of independence, as discussed  in Subsection~\ref{sub:dis3}, may be interpreted as the need for a conspiracy theory, or some version of superdeterminism, in order to achieve consistency with the Born rule. We tend to think of a measurement as something close to revealing a property of a system. In models of quantum measurement this independence of system and measurement apparatus is usually expressed by an initial state which is factorised between the system and the rest, and a hamiltonian which, up to a factorised interaction term, only consists on operators acting on either the system or the rest. For the reader that is attracted to this picture of independence, maintaining the problem of outcomes may be preferable to solving the problem of independence, as this perverts the idea of what a measurement is too much. The problem of epsilonics, see Subsection~\ref{sub:dis3}, can also be interpreted as an indication that the problem of outcomes cannot be solved within quantum mechanics. This is because the problem tells us that we cannot think of a state as a collection of properties. For the reader that still seeks solutions to the problem of outcomes along the lines of asymptotic Bohrification, the problem of epsilonics simply states that the Born rule must be assumed a priori.

If we dismiss counterfactual reasoning and arguments which depend on idealisations, then it is not a priori clear that the problem of outcomes cannot be solved within the formalism of unitary quantum theory. Unfortunately, the flea model does not provide us with the tools needed to fruitfully attack this problem.

\section{Acknowledgements}
The research in this paper is part of the project \emph{Experimental Tests of Quantum Reality}, funded by the \emph{Templeton World Charity Foundation}. The authors would like to thank Klaas Landsman for his investment in this project. We gratefully acknowledge the helpful discussions with Andrew Briggs, Hans Halvorson, Andrew Steane and various members of the Oxford Materials groups.

\appendix

\section{Appendix: Convergence of States} \label{sec:convergence}

For concreteness, let $\hbar$ take values in the unit interval $[0,1]$. To each strictly positive $\hbar>0$, associate the non-commutative algebra $\mathfrak{A}_{\hbar}=\mathcal{K}(L^{2}(\mathbb{R}))$ of compact operators acting on the Hilbert space of square-integrable functions. To $\hbar=0$ associate the commutative algebra $\mathfrak{A}_{0}=C_{0}(\mathbb{R}^{2})$ of continuous real-valued functions on the phase space $\mathbb{R}^{2}$, which vanish at infinity. Through their disjoint union $\mathfrak{A}=\coprod_{\hbar\in[0,1]}\mathfrak{A}_{\hbar}$ these algebras combine in a single algebra fibred over the unit interval, $\mathfrak{A}\to[0,1]$. Dual to this bundle there is the bundle of state spaces $\mathcal{S}\to[0,1]$ where $\mathcal{S}=\coprod_{\hbar\in[0,1]}\mathcal{S}_{\hbar}$, and $\mathcal{S}_{\hbar}$ denotes the state space of $\mathfrak{A}_{\hbar}$. For $\hbar>0$ the states are density operators acting on $L^{2}(\mathbb{R})$, and for $\hbar=0$ the states are probability measures on the phase space $\mathbb{R}^{2}$. 

Next, we could consider the algebraic and topological aspects of these bundles. But since we are only concerned with the problem of outcomes, we refer the reader to~\cite{landsman} and proceed directly to the our main question; how is convergence of states defined in this scheme? More precise, when does a family of density operators $(\rho_{\hbar})_{\hbar\in(0,1]}$ converge to a classical state $\mu_{0}\in\mathcal{S}_{0}$? To define convergence, note that each density operator $\rho_{\hbar}$ defines a probability measure $\mu_{\hbar}$ on $\mathbb{R}^{2}$, through
\begin{equation}
\int_{\mathbb{R}^{2}}d\mu_{\hbar}f:=Tr\left(\rho_{\hbar}Q_{\hbar}(f)\right),\ \ \ \forall f\in C_{0}(\mathbb{R}^2)
\end{equation}
where $Q_{\hbar}(f)$ is the compact operator acting on $L^{2}(\mathbb{R})$ defined to be the Berezin quantisation of $f$. The Berezin quantisation map $Q_{\hbar}$ is defined as
\begin{equation}
Q_{\hbar}(f)=\int_{\mathbb{R}^{2}}\frac{dp dq}{2\pi\hbar}f(p,q)\vert\Phi^{(p,q)}_{\hbar}\rangle\langle\Phi^{(p,q)}_{\hbar}\vert,
\end{equation}
through the coherent states $\Phi^{(p,q)}_{\hbar}\in L^{2}(\mathbb{R})$;
\begin{equation}
\Phi^{(p,q)}_{\hbar}(x)=(\pi\hbar)^{-1/4}e^{-ipq/2\hbar}e^{ipx/\hbar}e^{-(x-q)^{2}/2\hbar}.
\end{equation}  

The states $\rho_{\hbar}$ converge to the classical (possibly mixed) state $\mu_{0}$ iff the probability measures $\mu_{\hbar}$ converge weakly to $\mu_{0}$ in the sense
\begin{equation*}
\lim_{\hbar\to0}\int_{\mathbb{R}^{2}}d\mu_{\hbar}f=\int_{\mathbb{R}^{2}}d\mu_{0}f,
\end{equation*}
for each $f\in C_{0}(\mathbb{R}^{2})$ with compact support. For the double-well model, as $\hbar\to0$, the ground state $\psi_{0}(x)$, or rather its associated density operator, converges to the classical mixed state (\ref{equ_mixmaster}), a convex combination of probability distributions with support in the two different wells. A classical state which does not qualify as an outcome.

\section{A Matter of Necessity} \label{sec:nes}

By Earman's principle, mathematical idealisations should not bear physical significance. Following this principle, the problem of outcomes still seems a daunting challenge, but at least it is no longer an actual contradiction (unless we want to invoke counterfactual reasoning). Yet, in following the loopholes opened by Earman's principle, such as fuzzy eigenstate-eigenvalue links or approximate pointer bases, we find ourselves using approximate versions of outcomes. This reliance on approximations in asymptotic Bohrification is expressed by Butterfield's principle. Analogous  approximations arise in everettian quantum mechanics when invoking decoherence, and in the more realistic GRW models, which use gaussian collapsed states. If we indeed believe that an outcome is itself a classical idealisation, and that we should therefore ultimately understand measurement interactions in terms of the approximate counterparts, then we face the following problem: We want to use asymptotic Bohrification or one of the other approaches to solve the problem of outcomes  and the origin of the Born rule. Yet, in making these approaches understandable, by the very nature of the approximations involved, we find ourselves implicitly assuming the Born rule and unjustified certainties, in the form of pure states, from the start. In this section we show by example that the use of approximations is not a weakness of such approaches, but it adds necessary obstacles to be overcome in any approach which does not want to resort to classical approximations and the unjustified use of pure states. 

We concentrate on the example of preparing an initial system state using the Stern-Gerlach experiment. We find the same problem for the prepared system state with respect to the observable, as we previously did for the post-measurement state with respect to the pointer variable: For any \emph{fully} quantum mechanical treatment, the Born probabilities associated to both eigenvalues of the observable are always non-zero. If we started with an $n$-level system, the same would hold for all the eigenvalues of any observable. Consequently, this makes it hard to think of the system state as a collection of properties. The arguments used here, in particular our refusal to use classical approximations, could equally well be applied to a post-measurement state. The reason that we consider a state preparation, is that the entanglement of the prepared system state with the environment may bear relevance to the problem of independence. \\

Consider the Stern-Gerlach experiment where spin-$1/2$ particles
are sent through an inhomogeneous magnetic field. The textbook view
is that due to the spin-magnetic-field interaction the spin along
the magnetic field gets correlated with the position of the particle.
In this simple view of the device, the incoming particle is described
by a wavepacket $\psi(\mathbf{x},t)$ and then after some interaction
time the position of the wavepacket is correlated to the spin

\begin{align*}
\rbracket{\Psi(t=0)}= & \left(\alpha\rbracket{\uparrow}+\beta\rbracket{\downarrow}\right)\int\psi(\mathbf{x},0)\rbracket{\mathbf{x}}\mathrm{d}\mathbf{x}\\
\rightarrow\rbracket{\Psi(t)}= & \alpha\rbracket{\uparrow}\int\psi_{+}(\mathbf{x},t)\rbracket{\mathbf{x}}\mathrm{d}\mathbf{x}+\beta\rbracket{\downarrow}\int\psi_{-}(\mathbf{x},t)\rbracket{\mathbf{x}}\mathrm{d}\mathbf{x},
\end{align*}
where $\psi_{\pm}(\mathbf{x},t)$ indicate the wavepackets as they
leave the Stern-Gerlach apparatus. If the initial wavepacket is Gaussian,
the wavepackets $\psi_{\pm}$ will also be approximately Gaussian
but shifted upwards or downwards along the $z$-axis. For a derivation
of the typical form of such wavepackets see \cite{vdgrift, potel, roston}.

For the Stern-Gerlach apparatus to serve its purpose, to distinguish
spin states based on the position of the particle, the conditions
of the experiment should be such that most particles will be detected
at two well-separated positions on a detector screen. Some of the
dominant parameters, which determine the separation of the final wavepackets,
are the initial wavepacket width, strength and in-homogeneity of the
magnetic field, and the time of flight during and after the interaction. 

These parameters are varied by the experimenter when designing and
testing the experiment until the overlap between the wavepackets $\int\psi_{+}(\mathbf{x},t)$$\psi_{-}(\mathbf{x},t)\mathrm{d}t$
becomes extremely small such that for all practical purposes the wavepackets
seem to be separated in space. However, according to quantum theory
the overlap will in general always be non-zero. In other words, spin
is not perfectly correlated with position on the detector screen. 

If a small slit is made in the detector in the region we identify
with ``spin-up'', the state immediately after the slit will be given
by 
\[
\alpha\rbracket{\uparrow}\int\psi_{+}(\mathbf{x},t)\rbracket{\mathbf{x}}\mathrm{d}\mathbf{x}+\beta\rbracket{\downarrow}\int\psi_{-}(\mathbf{x},t)\rbracket{\mathbf{x}}\mathrm{d}\mathbf{x},
\]
where now the integration is restricted to the small slit. If the experiment
is well designed, one of the terms will be (exponentially) smaller
than the other. In principle, we should also allow for an extremely
small contribution where the particle tunnels through the detector
screen\footnote{Similarly, in the two-slit experiment a photon is said to travel through
both slits at once, however, in principle there is also a contribution
that it tunnels through the screen itself, which is dependent on the
thickness of the screen.}.

In practice, when the Stern-Gerlach apparatus is used as a preparation
device, the smaller term will be discarded as the parameters of the
setup were tuned specifically for reproducibility, i.e., it is tuned
such that the smaller term is experimentally inaccessible to subsequent
verification (using another device) due to the finite statistics and
the resolution of any experiment. This leads to the erroneous conclusion
that a pure state in spin-space can be obtained by application of
the Stern-Gerlach apparatus. Theoretically, after the slit the following
density matrix in the $\uparrow,\downarrow$-basis is obtained
\begin{equation}\label{eq:SG_dens}
\rho_{s}=\int\left(\begin{array}{cc}
\left|\alpha\psi_{+}\right|^{2} & \alpha^{*}\psi_{+}^{*}\beta\psi_{-}\\
\alpha\psi_{+}\beta^{*}\psi_{-}^{*} & \left|\beta\psi_{-}\right|^{2}
\end{array}\right)\mathrm{d}\mathbf{x}.
\end{equation}
Experimentally, the factors in the density matrix can be tuned more-or-less
continuously by the above-mentioned parameters, however, they will
never be strictly equal to one or zero unless the exact initial spin-state
was known, i.e., the exact value of $\alpha$ and $\beta$ is known
beforehand.

A further fundamental complication is that the magnetic field must
have zero divergence, which implies that it cannot have a gradient
in the field in only one direction \cite{potel}. Therefore,
as the wavepacket has a finite width in space, each part of it couples
to its local direction of the magnetic field which are not precisely
aligned with the single $z$-axis that is considered theoretically.
Thus particles with initially the same spin state along the quantization
axis can, nevertheless, deviate according to that of the opposite
spin.

\begin{figure}
\begin{centering}
\includegraphics[scale=0.25]{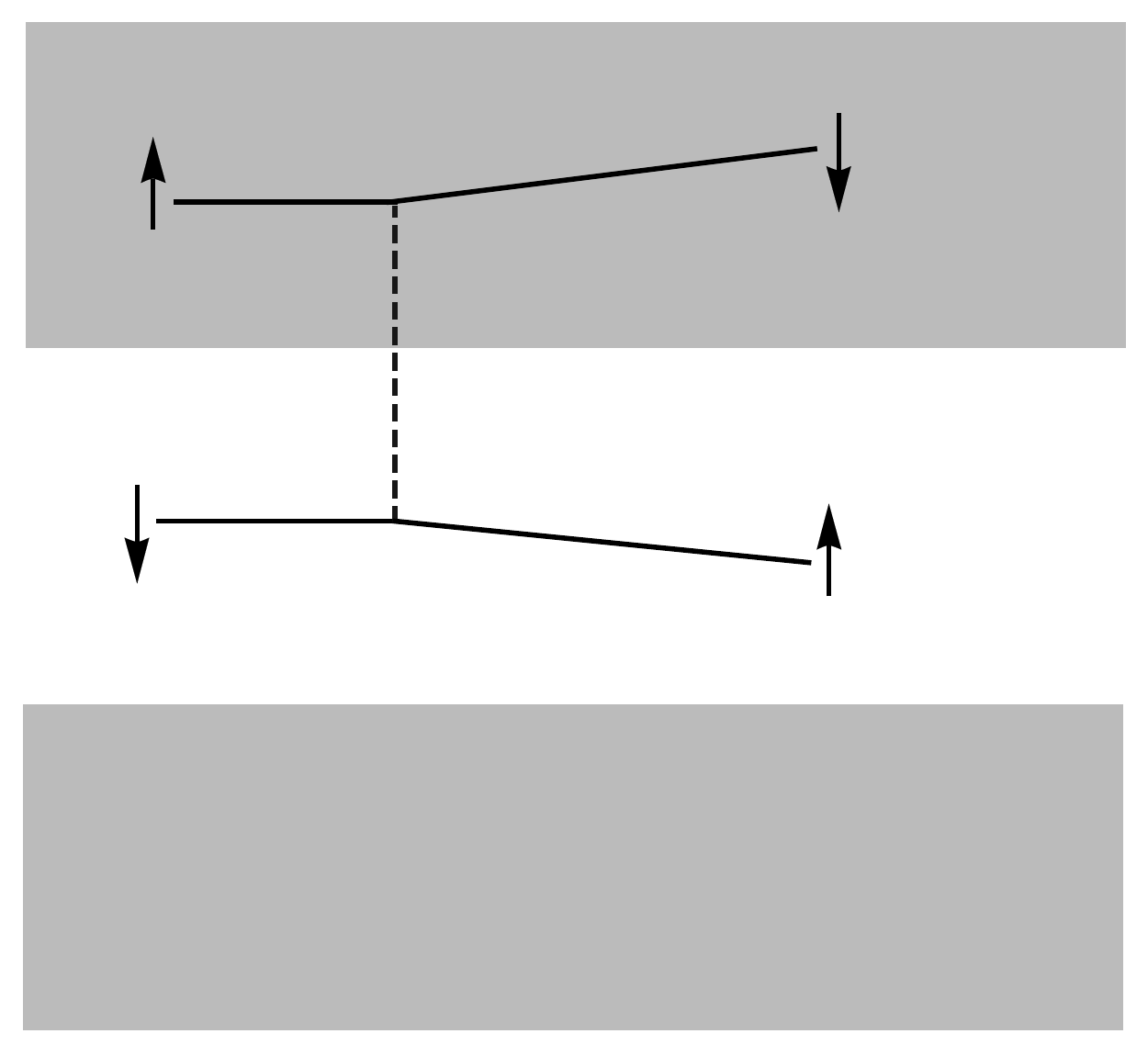}
\par\end{centering}

\protect\caption{A spin-flip process in the Stern-Gerlach experiment.\label{fig:A-spin-flip-process}}
\end{figure}

Another important point is that the magnetic field and magnet were
presumed to be classical. If the electromagnetic field is treated
quantum mechanically as mediating interactions between the spin-$1/2$
test-particle and the particles in the Stern-Gerlach magnets, it would
result in entanglement between the test-particle's position and those
of the charge carriers in the coils of the Stern-Gerlach magnets.
Namely, the charge carriers in the magnet would undergo a momentum
increase, and thereby change their position, along the $z$-direction
depending on the spin-component of the test-particle's wavefunction.
After tracing out the states of the magnet, a density matrix is obtained
similar to Equation (\ref{eq:SG_dens}). Also, as spin exchange processes between the test-particle
and the magnet's particles are always possible, there are again contributions
which cause incorrect deflections to occur. Such processes are easy
to visualize in the path-integral picture, which sums the amplitudes
over all possible paths and interactions, see Figure~\ref{fig:A-spin-flip-process}.
The suppression of spin-flips can be argued to be negligible under
typical circumstances due to Pauli blocking of transitions to already
occupied electronic states in the magnet, whereby we assume classical properties to the magnet.

Summarizing, the Stern-Gerlach experiment cannot be used as an ideal
and reliable preparation device of spin states, even in principle,
as there is no perfect one-to-one correspondence with position and spin. Note
that the objections to this experiment creating pure states in spin
are of a fundamental nature, namely they lie in the divergence-less
of the magnetic field, or the entanglement with the magnet which it
necessarily interacts, or the spatial extent of the wavefunctions. Furthermore, we meant to illustrate that when a pure state or a density matrix with some zeros on the diagonal is used, which has the benefit of being easy to interpret for single systems, it is in fact unjustified by the rules of quantum theory itself. We claim that the idealization of such states sneaks in a certainty about properties of single systems and can be found to be erroneous in any model by scrutinizing our classical assumptions and by thoroughly analysing sources of entanglement. In essence, it follows from the fact that all processes in quantum theory that can contribute will contribute, unless strictly forbidden by a conservation law\footnote{As an extreme example, suppose the proton is not absolutely stable. If it were only slightly unstable, this would mean that all measurement devices are always in a superposition. Restricting ourselves to only the non-decayed state results in a non-pure density matrix with all its interpretational issues.}. Following Earman's principle, we find that a truthful fundamental attempt at solving the measurement problem, which takes the rules of quantum theory seriously, must not rely on the artificial certainty of such states.

The impossibility of preparing a pure state, in the absence of classical approximations and entangling interactions, is not the only important point of this subsection. In Subsection~\ref{sub:born_baby_born} it was noted that the differences in initial environmental states (and the unitary operator) between different runs need to be linked to the Born probabilities associated to the observable of the system. In ignoring entanglement of the system with environmental degrees of freedom of the initial system state, we may lose the possibility of providing such a link between system and environment.

\end{document}